\documentclass[prb, superscriptaddress, twocolumn]{revtex4-1}
\usepackage{framed}
\usepackage{natbib}
\usepackage{graphicx}
\usepackage{dcolumn}
\usepackage{amsmath}
\usepackage{amssymb}
\usepackage{amsmath,verbatim,moreverb,bm}
\usepackage[normalsize,nooneline]{subfigure}
\usepackage{color}

\def\be{\begin{equation}}
\def\ee{\end{equation}}
\def\ber{\begin{eqnarray}}
\def\eer{\end{eqnarray}}

\def\sigmav{\mbox{\boldmath $\sigma$}}
\def\rv{{\bf r}}
\def\zv{{\bf \hat z}}
\def\pv{{\bf p}}

\def\Acalv{{\boldsymbol {\mathcal A}}}
\def\Acal{{\cal A}}

\begin{document}
\title{Room temperature spin thermoelectrics in metallic films}
\author{Sebastian T\"{o}lle}
\affiliation{Universit\"at Augsburg, Institut f\"ur Physik, 86135 Augsburg, Germany}
\author{Cosimo Gorini}
\affiliation{Universit\"at Augsburg, Institut f\"ur Physik, 86135 Augsburg, Germany}
\affiliation{Service de Physique de l'\'{E}tat Condens\'{e}, CNRS URA 2464, CEA Saclay, 91191 Gif-sur-Yvette, France}
\author{Ulrich Eckern}
\affiliation{Universit\"at Augsburg, Institut f\"ur Physik, 86135 Augsburg, Germany}


\begin{abstract}
Considering metallic films at room temperature, we present the first theoretical study of the spin Nernst and thermal Edelstein effects
which takes into account dynamical spin-orbit coupling, i.e., direct spin-orbit coupling with the vibrating lattice (phonons) and impurities.
This gives rise to two novel processes, namely a dynamical Elliott-Yafet spin relaxation and a dynamical side-jump mechanism.
Both are the high-temperature counterparts of the well-known ${T}=0$ Elliott-Yafet and side-jump, central to the current understanding
of the spin Hall, spin Nernst and Edelstein effects at low $T$.  We consider the experimentally relevant regime ${T}>{T}_D$,
with ${T}_D$ the Debye temperature, as the latter is lower than room temperature in transition metals such as Pt, Au and Ta typically employed
in spin injection/extraction experiments.  We show that the interplay between intrinsic (Bychkov-Rashba type)
and extrinsic (dynamical) spin-orbit coupling yields a nonlinear $T$-dependence of the spin Nernst and spin Hall conductivities. 
\end{abstract}
\maketitle

\section{Introduction}
\label{sec_intro}

Efficient heat-to-spin conversion is the central goal of spin caloritronics.\cite{bauer2012}
When considering metallic systems, two interesting phenomena stand out in this field:
the spin Nernst \cite{tauber2012,borge2013} and thermal Edelstein effects.\cite{pang2010,dyrdal2013}
They consist in the generation of, respectively, a spin current or a spin polarization
transverse to an applied temperature gradient.  That is, they are the thermal counterparts
of the well known spin Hall \cite{dyakonov1971} and Edelstein effects.\cite{aronov1989,edelstein1990}
These phenomena are due to spin-orbit coupling and do not require the presence 
of magnetic textures or Zeeman fields, and are typically classified as {\it intrinsic}
or {\it extrinsic} depending on their origin -- respectively band and device structure or impurities. 

Spin Hall measurements are typically performed in transition metals such as Au, Pt or Ta,\cite{vila2007,kimura2007, seki2008, liu2011, hahn2013}
where such effects are orders of magnitude larger than in standard semiconductors,\cite{kato2004}
and, very importantly, at room temperature.   In this temperature regime the dominant momentum-degrading
scattering mechanism in bulk is electron-phonon scattering.  Therefore the latter will, 
through spin-orbit coupling, heavily affect the spin Hall signals.  An identical reasoning applies to the Edelstein, thermal Edelstein and spin Nernst effects, 
though the last two have yet to be experimentally observed.  Indeed, the spin-orbit interaction adds an interesting twist to the coupling between electrons and phonons:
electrons in a disordered lattice at ${ T}=0$ move in a ``frozen'' electrostatic potential $U(\rv)=V_{\rm crys}(\rv)+V_{\rm imp}(\rv)$
arising from the crystal lattice and the impurities, yielding in the Hamiltonian the terms
\be
\label{intro1}
U(\rv) + \frac{\lambda_0^2}{4\hbar}\sigmav\times\nabla U(\rv)\cdot\pv,
\ee
$\sigmav,\pv$ and $\lambda_0$ being, respectively, the vector of Pauli matrices, the electron momentum and the Compton wavelength.
The potential becomes, however, time-dependent at finite $T$,
$U(\rv) \rightarrow U(\rv,t).$
Thus, the lattice (impurity) dynamics will not only give rise to standard electron-phonon (dynamical impurity) scattering
through the term $U(\rv,t)$, but will also couple directly to the carrier spin through the {\it dynamical} spin-orbit interaction 
$\lambda_0^2\sigmav\times\nabla U(\rv,t)\cdot\pv/4\hbar$.
Remarkably, such a direct ``spin-phonon'' (``spin-dynamical impurities'') coupling has not yet been studied, and even standard electron-phonon scattering
has received minimal attention in the present context.  To the best of our knowledge, the only theoretical work considering the impact of standard electron-phonon interaction 
on the spin Hall effect is that of Grimaldi et al.~\cite{cappelluti2006}, which is focused on a 2-Dimensional Electron Gas (2DEG) with Bychkov-Rashba\cite{bychkov1984} spin-orbit coupling at $T\rightarrow0$.

Our purpose is to start filling this gap, considering the spin Hall, spin Nernst, Edelstein and thermal Edelstein effects in a metallic thin film at room temperature.
Moreover, we wish to identify the possible connections between the four phenomena.
It is known, for example, that in a 2DEG at low $T$ the spin Hall and Edelstein effects
are closely related,\cite{gorini2008,raimondi2012} and that such a relation can be extended to thin
(quasi-2D) films as well.\cite{borge2014}  Whether this connection exists, possibly in a modified form, also at high $T$ or in 3D is an open question.  
Another important point concerns the $T$-dependence of the above cited effects.
For example, whereas this is expected to be linear for a purely extrinsic 
spin Nernst effect \cite{tauber2012}, 
it is not known how the interplay between extrinsic and intrinsic mechanism will modify such behavior.
Similarly, for the spin Hall effect its $T$-dependence should allow to establish its specific intrinsic or extrinsic origin.\cite{vignale2010,niimi2011}  The latter is still a somewhat
controversial issue, in particular in Au and Pt.\cite{vila2007,niimi2011,niimi2014,isasa2014}

Our treatment relies on two central assumptions. 
The first one is based on the observation that the Debye temperature ${T}_D$ of bulk Au (165 K), Pt (240 K) or Ta (240 K) is lower than
room temperature, and in this regime electron-phonon scattering is predominantly {\it elastic}.\cite{zimanbook1} 
This leads to a remarkable simplification of the quantum kinetic
equations we will employ, allowing to extend to the present case the analysis of the $T=0$ scenario.\cite{zimanbook1} 
The second one concerns the type of spin-orbit interaction in a metallic film.  There is yet no general theory capable of identifying its precise effective form,
but experiments show that a strong Rashba-like spin-orbit interaction
appears at the interface between transition metals and insulators/vacuum, \cite{shikin2008,varykhalov2008,rybkin2010} where inversion symmetry is broken; density functional theory has been recently used to estimate its size in Ag, Au or Al on W(110) structures.\cite{hortamani2012}
In general, spin-orbit splittings of up to several hundreds of meV are reported -- that is, considerably larger than in a standard GaAs 2DEG.
We will thus assume the intrinsic spin-orbit mechanism to be described by a Rashba term in the Hamiltonian.
The extrinsic one will be treated in analogy with the semiconductor case, where the spin-orbit interaction with the impurity potential is
mediated by an effective Compton wavelength renormalized by the lattice.\cite{winklerbook,Handbook}

Experimentally realized films explore the full 2D to 3D range, thicknesses ranging from
one or few monolayers \cite{shikin2008,varykhalov2008,rybkin2010}, up to few to tens of nanometers. \cite{valenzuela2006,seki2008,hahn2013}  
We will start by considering a strictly 2D metallic layer, and later comment on its 3D counterpart.
For the latter case our approach follows the spirit of Ref.~[\onlinecite{pesin2012}], which takes the Rashba-like term to be {\it homogeneous} across the film thickness.  
Notice that this is complementary to what is done in Refs.~[\onlinecite{wang2013,haney2013,borge2014}], where the Rashba interaction is a $\delta$-function different from zero 
only {\it exactly} at the film edges.

Finally, we will rely on the $SU(2)$-covariant kinetic formulation introduced in Ref.~[\onlinecite{gorini2010}].  
This considerably simplifies the collision integrals to be faced,\cite{gorini2010,raimondi2012} and ensures the unambiguous definition of spin-related physical
quantities even when the spin itself is not conserved (due to spin-orbit interaction).\cite{tokatly2008,gorini2010}
In particular, as shown in Ref.~[\onlinecite{gorini2012}], it provides the framework 
to properly define Onsager reciprocal relations in the presence of spin-orbit coupling,
e.g., between the direct and inverse spin Hall\cite{gorini2012} or Edelstein\cite{shen2014} effects.  
This ensures that our results will have an immediate bearing on the inverse counterparts of the phenomena considered below.

The paper is organized as follows.  We first introduce the model and the linear response
formulation in Sec.~\ref{sec_model}, then move on to the kinetic approach in Sec.~\ref{sec_kinetic}.
Section~\ref{sec_effects} discusses the core results, namely the spin Nernst and thermal
Edelstein effects.  The focus is on their $T$-dependence and their relation with each other, 
as well as with the spin Hall and Edelstein effects.  We conclude with a brief summary.
Certain general but cumbersome formulas are given in Appendix \ref{App},
whereas the estimation of different spin lifetimes appear in Appendix \ref{App_taus}. 

\section{The model and the Onsager formulation}
\label{sec_model}  

Let us start from the following effective (static) model Hamiltonian for conduction electrons in a parabolic band:
\footnote{Finer band structure details such as non-parabolicity can be included in the kinetic treatment to follow 
(Ref.~[\onlinecite{shytov2006}], though concerned with a semiconductor scenario, gives a taste of the technicalities involved.).  
However, since they are not central to our goals, we stick to a bare-bone model for simplicity's sake.}
\be
\label{model1}
H_0 = \frac{p^2}{2m} - \frac{\alpha}{\hbar}\sigmav\times\hat{\bf z}\cdot\pv + V_{\rm imp}(\rv) -
\frac{\lambda^2}{4\hbar}\sigmav\times\nabla V_{\rm imp}(\rv)\cdot\pv.
\ee
As customary, the static lattice potential $V_{\rm crys}(\rv)$ does not appear explicitly anymore, its effects having been incorporated
in the effective mass $(m_0\rightarrow m)$ and effective Compton wavelength $(\lambda_0\rightarrow\lambda)$.\cite{winklerbook,Handbook} 
Above, $\zv$ is the unit vector pointing towards the metal-substrate interface, whereas $\pv, \rv$ can be either
vectors in the $x$-$y$ plane for strictly 2D films, or also have a $z$-component for thicker, 3D systems. 
The second term on the r.h.s.\ is the Bychkov-Rashba intrinsic spin-orbit coupling due to structure symmetry breaking (metal-substrate interface),
characterized by a coupling constant $\alpha$, whose strength can be measured by angle-resolved photoemission,\cite{shikin2008,varykhalov2008,rybkin2010} 
and estimated by ab-inito methods.\cite{hortamani2012}
$V_{\rm imp}(\rv)$ is the random impurity potential, see Sec.~\ref{sec_kinetic}.
Impurities give also rise to the fourth term, which represents extrinsic spin-orbit interaction.  
In the strictly 2D limit the Hamiltonian \eqref{model1} was used to study the spin Hall\cite{hankiewicz2008,raimondi2009,raimondi2012,gorini2012} and Edelstein effect\cite{raimondi2012} in the presence of both intrinsic and extrinsic mechanisms at $T=0$.  Such mechanisms were shown {\it not} to be simply additive, and their interplay leads to a nontrivial behavior.\cite{raimondi2009,raimondi2012}

For finite temperatures (${T}\neq0$) the now time-dependent potential $U(\rv,t)$ is expanded around its static configuration:
\ber \label{fullpotential}
U(\rv,t) &=& V_{\rm imp}(\rv) + \delta V_{\rm crys}(\rv,t) + \delta V_{\rm imp}(\rv,t) + \dots ,
\eer
where $\delta V_{\rm crys}(\rv,t),\,\delta V_{\rm imp}(\rv,t)$ are linear in the small ion/impurity displacements.
Note that the static lattice potential $V_{\rm crys}(\rv)$ has already been effectively taken into account, and so it does not appear in Eq.~(\ref{fullpotential}) above.  Neither does the phononic term, since we are not
interested in the phonon dynamics; the phonons are assumed to be in equilibrium.  The Hamiltonian thus becomes
\ber
\label{model2}
H &=& H_0 + \delta V_{\rm crys}(\rv,t) + \delta V_{\rm imp}(\rv,t) \nonumber \\
&&-\frac{\lambda^2}{4\hbar}\sigmav\times\nabla \left[\delta V_{\rm crys}(\rv,t) + \delta V_{\rm imp}(\rv,t)\right]\cdot\pv.
\eer
The second term on the r.h.s.\ gives rise to the electron-phonon interaction,
the third to electron scattering with dynamical impurities, and the fourth describes dynamical spin-orbit coupling (see Fig.~\ref{Diagrams}).  This last one is novel and crucial for our purposes,
as it yields the dynamical Elliott-Yafet spin relaxation and the dynamical side-jump mechanism.
Neither of these two processes have been considered previously, even though their
static counterparts are central in $T=0$ treatments of the spin Hall and related effects.
\cite{sinitsyn2008,culcer2010,raimondi2012}
A third potentially relevant process is phonon skew scattering.\cite{isasa2014}
This will be discussed elsewhere,\cite{skewfuture} since its treatment requires going beyond the Born approximation, which is beyond the scope of the present work.

In order to employ the $SU(2)$-covariant kinetic formulation\cite{gorini2010} mentioned in the Introduction,
the intrinsic Bychkov-Rashba term is rewritten as a non-Abelian vector potential:\cite{mathur1992,frohlich1993,tokatly2008,gorini2010}
\be
- \frac{\alpha}{\hbar} p_i \varepsilon_{iaz} \sigma^a = \frac{p_i\Acal_i^a\sigma^a}{2m} ,
\ee
with $\Acal^x_y=-\Acal^y_x=2m\alpha/\hbar$, all other components of $\Acalv^a$ being zero, whereas $\varepsilon_{iaz}$ is the $z$-component of the antisymmetric tensor.  
Here and throughout the paper upper (lower) indices will indicate spin (real space) components,
while repeated indices are summed over unless otherwise specified.  

The final step is defining the relevant transport coefficients within linear response.
Assuming homogeneous conditions and taking as driving fields an electric field $E_x$ and a temperature gradient $\nabla_xT$,
we are interested in the generation of (i) a $y$-spin polarization $s^y$ (Edelstein\cite{aronov1989,edelstein1990} and thermal Edelstein\cite{pang2010,dyrdal2013} effects);
(ii) a $z$-polarized spin current flowing along $y$, $j^z_y$ (spin Hall\cite{dyakonov1971} and spin Nernst\cite{tauber2012,borge2013} effects).
In the presence of spin-orbit coupling, i.e., when spin is not conserved, the spin current has a diffusion term
even under homogeneous conditions:\cite{gorini2010}
\be
\label{scurrent1}
j^z_y = 2m\alpha D s^y + j^z_{y, {\rm drift}},
\ee  
with $D$ the diffusion constant.
Extending the standard Onsager formulation of thermoelectric transport to the present spin-thermoelectric context, we then write
\ber
\label{edelstein1}
s^y &=& P_{\rm sE} \,E_x + P_{\rm sT} \,\nabla_x T ,
\\
\label{drift1}
j^z_{y, {\rm drift}} &=& \sigma_{\rm sE, drift} \,E_x + \sigma_{\rm sT, drift} \,\nabla_x T.
\eer
The conductivities $\sigma_{\rm sE, drift}, \sigma_{\rm sT, drift}$ correspond, in Kubo diagrammatics, to ``bare'' response bubbles.
For the full spin current $j^z_y$ one has
\be
\label{scurrent2}
j^z_y = \sigma_{\rm sE} \,E_x + \sigma_{\rm sT} \,\nabla_x T.
\ee
where $\sigma_{\rm sE}, \sigma_{\rm sT}$ are bubbles with ``dressed'' vertices, the same holding for $P_{\rm sE}, P_{\rm sT}$.  
The spin Hall conductivity $\sigma^{\rm sH}\equiv\sigma_{\rm sE}$,
whereas the spin Nernst one is defined under open circuit conditions, $\sigma^{\rm sN}\equiv S\sigma_{\rm sE}+\sigma_{\rm sT}$, with $S$ the Seebeck coefficient.
Similarly, the Edelstein effect is directly given by the spin polarization response to the electric field, ${\mathcal P}\equiv P_{\rm sE}$, while for its thermal counterpart
${\mathcal P}^{t}\equiv SP_{\rm sE}+P_{\rm sT}$.

Our goal is the computation of the transport coefficients $P_{\rm sE}, P_{\rm sT}, \sigma_{\rm sE}, \sigma_{\rm sT}$ defined above.  For the sake of clarity we have introduced them within a drift-diffusion picture,
however Eqs.~\eqref{edelstein1} and \eqref{scurrent2} are general, and our treatment 
works in the ballistic limit as well.
Finally, Onsager reciprocity is duly respected,\cite{gorini2012,shen2014}
and is here between $j^z_y\leftrightarrow j_x$ (spin Hall $\leftrightarrow$ inverse spin Hall effect) 
and $s^y\leftrightarrow j_x$ (Edelstein $\leftrightarrow$ inverse Edelstein or spin-galvanic effect\cite{ganichev2002,rojassanchez2013}).

\section{The kinetic equations}
\label{sec_kinetic}
\def\Fcalv{{\boldsymbol {\mathcal F}}}

The kinetic (Boltzmann-like) equation for the $2\times2$ distribution function $f_{\pv}=f^0+\sigmav \cdot {\bf f}$, where $f^0$ and $\bf f$ are the charge and spin distribution functions, respectively,\cite{gorini2010} reads
\be \label{Boltzmann}
\partial_t f_{\pv} + \tilde{\nabla} \cdot \left[ \frac{\pv}{m} f_{\pv} + \Delta {\bm j}_{\rm sj} \right] + \frac{1}{2} \left\lbrace{\Fcalv}\cdot{\nabla}_{\pv} , f_{\pv} \right\rbrace= I_{0}+I_{\rm sj}+I_{\rm EY} ,
\ee
where we introduced the covariant spatial derivative and the $SU(2)$ Lorentz force due to the Rashba spin-orbit coupling:
\ber
\tilde{\nabla}{}&{}={}&{}\nabla+\frac{i}{\hbar}\left[ \Acalv^a\frac{\sigma^a}{2},\cdot \right],\\
\Fcalv{}&{}={}&{}- \frac{\pv}{m} \times \boldsymbol{\mathcal{B}}^a \frac{\sigma^a}{2},\\
\mathcal{B}_i^a{}&{}={}&{}-\frac{1}{2\hbar}\varepsilon_{ijk} \varepsilon^{abc}\mathcal{A}_j^b\mathcal{A}_k^c .
\eer
A summation over identical indices is implied unless stated otherwise. Note that an external magnetic field is not included in these equations (since it is not needed for the present purpose). The term $\Delta {\bm j}_{\rm sj}$ in Eq.~\eqref{Boltzmann} is a correction to the current due to side-jumps. 

\begin{figure}
\includegraphics[width=8cm]{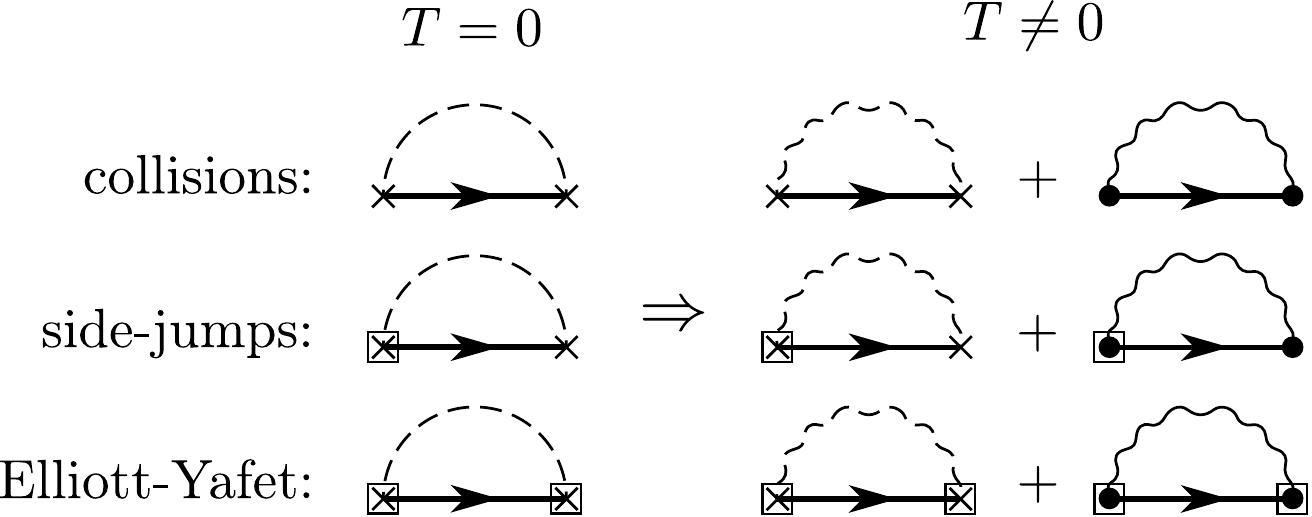}
\caption{Shown are the self-energies which determine the collision operators in the Boltzmann equation. The arrowed line represents the Green's function in Keldysh space, a cross (dot) the potential due to an impurity (a crystal displacement). The dashed line depicts the impurity correlation either for static (straight line) or for dynamical impurities (wavy line). The wavy solid line illustrates the phonon propagator and a box around a vertex the spin-orbit coupling due to the boxed potential. \label{Diagrams}}
\end{figure}

Next we consider the collision operators on the r.h.s.\ of Eq.~\eqref{Boltzmann}, where $I_{0}$ describes scattering with dynamical impurities and phonons, $I_{\rm sj}$ the contribution due to side-jumps, and $I_{\rm EY}$ Elliott-Yafet spin relaxation due to spin-flip processes.  
At zero temperature the collision operators are obtained from the impurity averaged self-energies within the self-consistent Born approximation (see Fig.~\ref{Diagrams}). For isotropic scattering, the impurity correlations are given by
\be \label{impcorrelation1}
\overline{ V_{\rm imp}(\rv) V_{\rm imp}(\rv') }= n_{\rm imp} v_0^2 \delta(\rv-\rv') =\frac{\hbar}{2\pi N_0 \tau_{\rm imp}}\delta(\rv-\rv') ,
\ee
with $n_{\rm imp}$ the impurity concentration, $v_0$ the scattering amplitude, and $1/\tau_{\rm imp}$ the momentum relaxation rate due to impurities; $N_0$ is the density of states per area (volume) and spin in two (three) dimensions. More generally, $v_0^2 \rightarrow \langle|v({\bf q})|^2\rangle$, where $\langle\dots\rangle$ denotes the angular average, and ${\bf q}^2=(\pv-\pv')^2/\hbar^2=2p_F^2(1-\cos \theta)$, since $|\pv|=|\pv'|=p_F$.

In order to include the impurities' thermal fluctuations, we consider small time-dependent displacements $\delta \rv_i(t)$ of the $i$-th impurity, which leads to
\be
\delta V_{\rm imp}({\bf r},t)=-\nabla\cdot \sum\limits_i  \delta \rv_i(t) v(\rv-\rv_i),
\ee
where $v$ is the single-impurity potential. We further assume that the displacement fluctuations of different impurities are independent, and can be approximated by the classical harmonic oscillator expression, i.e,
\be
\overline{\delta r_i^\alpha(t)\delta r_j^\beta(t')} \simeq \delta_{ij}\delta_{\alpha \beta} \frac{k_B T}{M\omega_D^2},
\ee
where $M$ and $\omega_D$ are the typical mass and frequency; we also considered short times, $\omega_D |t-t'| \ll 1$. Then we obtain
\be \label{dyncorrelation}
\overline{\delta V_{\rm imp}(\rv,t)\delta V_{\rm imp}(\rv',t')} \simeq \frac{\hbar}{2\pi N_0 \tau_{\rm dyn}} \delta(\rv-\rv')
\ee
with
\be \label{taudyn}
\frac{1}{\tau_{\rm dyn}} = \frac{2\pi n_{\rm imp} v_0^2 N_0}{\hbar} \frac{2 k_B T p_F^2}{\hbar^2 M \omega_D^2}.
\ee
More precisely, as follows from the corresponding self-energy expression (Fig.~\ref{Diagrams}), $v_0^2\rightarrow \langle (1-\cos \theta) |v({\bf q})|^2 \rangle$ in \eqref{taudyn}. In order of magnitude, $\tau_{\rm imp}/\tau_{\rm dyn} \simeq k_B T / \epsilon_F$ since $(\hbar \omega_D)^2 \simeq (m/M) \epsilon_F^2$. Note that the $\delta$-function in Eq.~(\ref{dyncorrelation}) has to be interpreted in connection with the corresponding self-energy diagram. A detailed analysis shows that the result given in Eq.~(\ref{taudyn}) applies for high temperatures, $k_B T \gg \hbar \omega_D$, where scattering processes essentially are elastic. 

A similar reasoning can be employed for electron-phonon scattering at high $T$, which leads to
\be
\overline{ \delta V_{\rm crys}(\rv,t) \delta V_{\rm crys}(\rv',t')} \simeq \frac{\hbar}{2\pi N_0 \tau_{\rm ph}}\delta(\rv-\rv') ,
\ee
where $1/\tau_{\rm ph} = 2\pi N_0 g^2 k_B T/\hbar$ is the standard (high $T$) momentum relaxation rate.\cite{rammer} Based on the Keldysh technique, the collision operators can be derived as usual.\cite{rammer2} The result corresponds, in the classical limit, to
\be
\overline{\delta V_{\rm crys}(\rv,t)\delta V_{\rm crys}(\rv',t')}=\frac{i g^2}{2} D^K(\rv-\rv',t-t') , 
\ee
where $g$ is the electron-phonon coupling constant and $D^K$ denotes the Keldysh component of the phonon Green's function in equilibrium.

Since $1/\tau_{\rm ph}$ can be several orders of magnitude larger than $1/\tau_{\rm imp}$,\cite{ashcroft} the total momentum relaxation rate $1/\tau=1/\tau_{\rm imp}+1/\tau_{\rm dyn}+1/\tau_{\rm ph}$ is typically dominated by electron-phonon scattering, $1/\tau\simeq 1/\tau_{\rm ph}$, in the high-temperature regime. 

The above discussion shows that one may use the results for the collision operators and the side-jump correction given in Refs.~[\onlinecite{raimondi2012}] and [\onlinecite{gorini2010}],
\ber
I_{0} {}&{} = {}&{} -\frac{1}{\tau}(f_{\pv}-\langle f_{\pv} \rangle), \label{EPop} \\
I_{\rm sj}{}&{} = {}&{} \frac{\lambda^2}{8\hbar\tau}  \varepsilon_{abc} \left\{ (\tilde{\nabla}_a \sigma^b), p_c f_{\pv} -\left\langle p_c f_{\pv}\right\rangle \right\}  , \label{SJop} \\
I_{\rm EY}{}&{} = {}&{}-\frac{1}{\tau}\left(\frac{d-1}{d} \right)  \left( \frac{\lambda p}{2\hbar} \right)^4 \nonumber \\
{}&{}&{} \times \sum\limits_{a=x,y,(z)} \left( \frac{1}{3^{d-2}} f^a + \langle f^a \rangle  \right) \sigma^a \label{EYop} \\
\Delta {\bm j}_{\rm sj} {}&{} = {}&{} \frac{\lambda^2}{8\hbar \tau} \left\langle \left\{ \left( \pv'-\pv \right) \times \sigmav , f_{\pv'} \right\} \right\rangle_{\hat{\pv}'} , \label{SJcorr}
\eer
where $1/\tau$ is now the total scattering rate. The wavy brackets represent the anti-commutator and $d=2,3$ the dimensionality.\footnote{We remark that in three dimensions Eq.~(\ref{EYop}) is not the complete collision operator as obtained from the self-energy depicted in Fig.~\ref{Diagrams}: we have dropped terms where the momentum is not parallel to the Pauli vector. These contributions are negligible when investigating the spin transport quantities.} 
Formally, the diagrams in Fig.~\ref{Diagrams}, together with Eqs.~\eqref{EPop}-\eqref{EYop}, show that the phenomenological substitution $1/\tau_{\rm imp}\rightarrow1/\tau$ for $T=0\rightarrow T\neq0$ is fully justified for all spin-dependent processes at the Born approximation level of accuracy.

Finally, the $y$ spin polarization $s^y$ and the $z$-polarized spin current flowing along $y$, $j^z_y$, are defined according to Ref.~[\onlinecite{raimondi2012}], 
\be \label{spinPol}
s^y=\int \frac{\mathrm{d} \pv}{(2\pi\hbar)^d} f^y = \int \mathrm{d} \epsilon_{\pv} N_0 \langle f^y \rangle ,
\ee
and
\be \label{defspincurrent}
j_y^z=\mathrm{Tr}\frac{\sigma^z}{2} \int \frac{\mathrm{d} \pv}{(2\pi\hbar)^d}\left[\frac{p_y}{m} f_{\pv}+\frac{\lambda}{8\hbar\tau} \left\{ (\pv \times \sigmav) \right\}_y , f_{\pv} \right] .
\ee

%

\section{Spin Nernst and thermal Edelstein effects}
\label{sec_effects}
In this section we present and discuss our results, i.e., the spin transport coefficients $P_{\rm sE}, P_{\rm sT}, \sigma_{\rm sE}$, and $\sigma_{\rm sT}$.  We find that the competition between intrinsic
and extrinsic spin-orbit mechanisms can lead the former to have a nonlinear temperature dependence.
Notice that when only extrinsic mechanisms are considered, the spin Nernst conductivity 
was instead predicted to be simply linear in $T$.\cite{tauber2012}
Though the spin Nernst nonlinearity will prove to be rather weak in a wide range of parameters,
it is in principle a signature of the relative strength between intrinsic and extrinsic spin-orbit coupling.

We first consider a two-dimensional system and comment on the three-dimensional case at the end of this section.
Furthermore, we focus on the diffusive (``dirty'') regime, in which a very transparent drift-diffusion picture
for both charge and spin degrees of freedom is possible.\cite{gorini2010}
However, the ballistic (``clean'') limit is also discussed in the closing Subsection \ref{subsec_clean}, since estimates show it to be relevant for certain experimentally realized systems.  Indeed, spin diffusion takes place as long as the spin-orbit splitting is smaller than the lifetime broadening, which in a Rashba-like system means 
$2\alpha p_F/\hbar<\hbar/\tau$, $p_F$ being the Fermi surface momentum.  
At room temperature $\hbar/\tau\approx10^{-2}$eV, whereas $2\alpha p_F/\hbar$ can vary
substantially in metallic films, $10^{-3}{\rm eV}\lesssim2\alpha p_F/\hbar\lesssim10^{-1}{\rm eV}$.\cite{shikin2008,
varykhalov2008,rybkin2010}  Thus, the full diffusive-to-ballistic spectrum can in principle be explored.

In the diffusive regime the Boltzmann equation (\ref{Boltzmann}) for $\langle f^y \rangle$ can be solved within the $p$-wave approximation ($f_{\pv}\simeq \langle f_{\pv} \rangle + \hat{\pv}\cdot\delta {\bf f}_{\pv}$),
in terms of the $x$-spatial derivative of the local equilibrium charge distribution function,
\be
\nabla_x f^{eq}=  \left( \frac{\epsilon_{\pv}-\epsilon_F}{T} \nabla_x T + \nabla_x \mu \right) \left(- \frac{\partial f^{eq}}{\partial \epsilon_{\pv}} \right) .
\ee
Here $\epsilon_{\pv} (\epsilon_F)$ is the particle (Fermi) energy. The chemical potential gradient is identified with the electric field, $eE_x \equiv \nabla_x \mu$ with $e=|e|$. The temperature gradient and the electric field act as driving terms in the charge sector of the Boltzmann equation, which is easily solved. Via Eqs.~\eqref{SJop} and \eqref{SJcorr}, the charge distribution enters the spin sector, from which we determine $\langle f^y \rangle$ and hence the spin polarization linear in $E_x$ and $\nabla_x T$ according to Eq.~\eqref{spinPol}. In the last step, integrating the $y$ spin component of Eq.~\eqref{Boltzmann} we obtain
\be \label{conteq}
\partial_t s^y + \frac{2m\alpha}{\hbar^2} j_y^z = - \int \mathrm{d} \epsilon_{\pv} \frac{N_0}{\tau_s} \langle f^y \rangle .
\ee
From this relation, we then calculate $j_y^z$. Note that no spatial gradients (beyond $\nabla_x T$ and $\nabla_x \mu$) are considered. In Eq.~\eqref{conteq}, the (weakly energy-dependent) Elliott-Yafet relaxation rate is proportional to the momentum relaxation rate, and given by
\be
\frac{1}{\tau_s}=\frac{1}{\tau} \left( \frac{\lambda p}{2\hbar} \right)^4 .
\ee
Specifically, in order to obtain $\langle f^y \rangle$ we perform a Fourier transformation in time, $t\rightarrow \omega$, multiply the $z$ spin component of the Boltzmann equation by $p_y$, the charge component by $p_x$, and perform the momentum angular average of these two equations as well as of the $y$ spin component of the Boltzmann equation. The result is
\be
\langle f^y \rangle = - F_{\omega} \cdot \nabla_x f^{eq} 
\ee
with
\ber
F_{\omega}{}&{}={}&{} p^2 \frac{\alpha}{\hbar^3} \frac{\tau_s}{1-i\omega\tau} \left[ 2 \left( \frac{\alpha \tau}{\hbar} \right)^2 + \frac{\lambda^2}{2}(1-i\omega \tau) \right] \nonumber \\
{}&{}\phantom{=}{}&{}\times\left[ 2 \left( \frac{4\alpha\tau}{\lambda^2 p} \right)^2 + (1-i\omega\tau_s)(1-i\omega\tau) \right]^{-1} .
\eer
From this expression, we are now able to determine the transport coefficients, similar to Mott's formula in thermoelectrics.\cite{Mott1969} We find
\ber
P_{\rm sE}(\omega){}&{}={}&{}-e\int \mathrm{d} \epsilon_{\pv} N_0 F_\omega \left(- \frac{\partial f^{eq}}{\partial \epsilon_{\pv}} \right) , \label{Mott1} \\
P_{\rm sT}(\omega){}&{}={}&{}-\int \mathrm{d} \epsilon_{\pv} N_0 F_\omega \frac{\epsilon_{\pv}-\epsilon_F}{T} \left(- \frac{\partial f^{eq}}{\partial \epsilon_{\pv}} \right) , \\
\sigma_{\rm sE}(\omega){}&{}={}&{}\frac{e\hbar^2}{2m\alpha}\int \mathrm{d} \epsilon_{\pv} \frac{N_0}{\tau_s} \left(1-i\omega\tau_s \right) F_\omega \left(- \frac{\partial f^{eq}}{\partial \epsilon_{\pv}} \right) ,\hspace*{15pt} \\
\sigma_{\rm sT}(\omega){}&{}={}&{}\frac{\hbar^2}{2m\alpha}\int \mathrm{d} \epsilon_{\pv} \frac{N_0}{\tau_s} \left(1-i\omega\tau_s \right) F_\omega \nonumber \\
{}&{}&{}\hspace*{50pt}\times\frac{\epsilon_{\pv}-\epsilon_F}{T} \left(- \frac{\partial f^{eq}}{\partial \epsilon_{\pv}} \right) . \label{Mott2}
\eer
In the following we consider the first non-vanishing order of the Sommerfeld expansion\cite{ashcroft} of Eqs.~(\ref{Mott1})--(\ref{Mott2}). Also, all energy-dependent quantities are given at the Fermi energy unless mentioned otherwise. The following Mott-like formulas are obtained:
\ber
P_{\rm sT}{}&{}={}&{}-S_0 \epsilon_F P'_{\rm sE} , \label{Mott3} \\
\sigma_{\rm sT}{}&{}={}&{}-S_0\epsilon_F \sigma'_{\rm sE} , \label{Mott4}
\eer
with $S_0=-\pi^2 k_B^2T/(3e\epsilon_F)$, $P'_{\rm sE}\equiv\partial_{\epsilon_{\pv}}P_{\rm sE}|_{\epsilon_F}$, and $\sigma'_{\rm sE}\equiv\partial_{\epsilon_{\pv}}\sigma_{\rm sE}|_{\epsilon_F}$.

\begin{figure}
\flushleft{a)}\\
\includegraphics{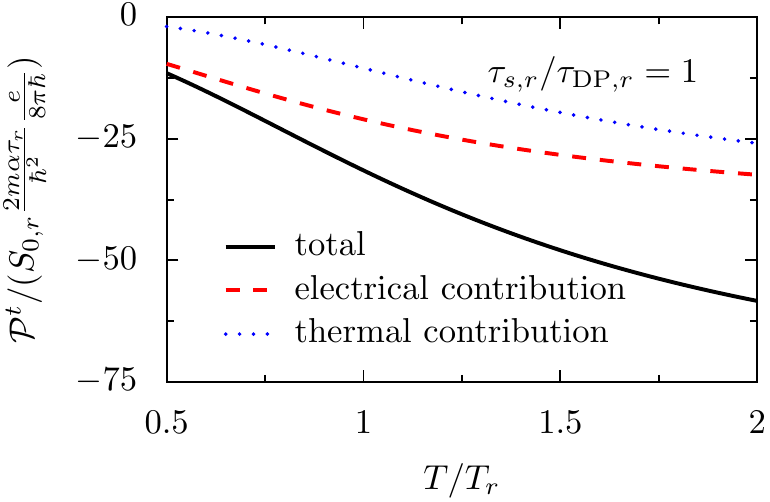}
\flushleft{b)}\\
\includegraphics{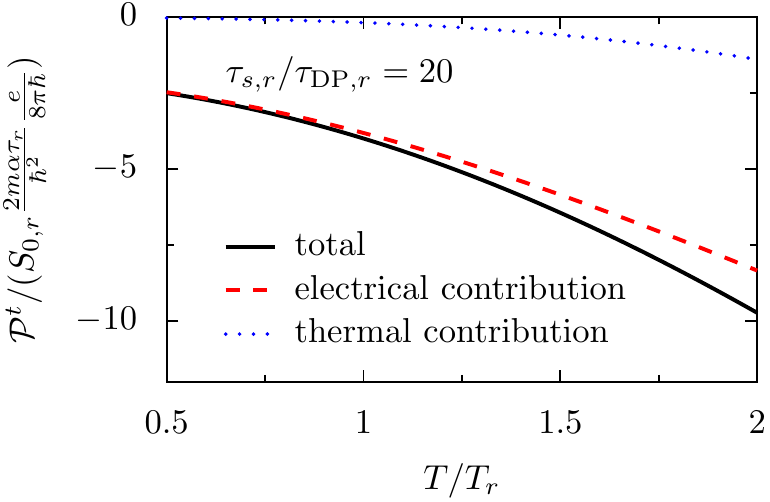}
\caption{${\cal P}^t$, compare Eq.~\eqref{PtEdef}, versus temperature, in units of $S_{0,r}(2m\alpha\tau_r/\hbar^2)(e/8\pi\hbar)$, split into its thermal and electrical contributions. The Elliottt-Yafet spin relaxation is chosen as $\tau/\tau_s=0.01$. For a) we have $\tau_{s,r}/\tau_{{\rm DP},r}=1$ and for b) $\tau_{s,r}/\tau_{{\rm DP},r}=20$. $T_r$ denotes the temperature scale (room temperature). \label{PtE}}
\end{figure}

First we discuss the simple case of a 2DEG with an energy-independent relaxation rate $1/\tau$ in the static case ($\omega=0$). We refer to App. \ref{App} for more general formulas. Concerning the spin polarization, we find
\ber
P_{\rm sE}{}&{}={}&{}-\frac{2m\alpha}{\hbar^2} \frac{\tau_s}{\tau_s/\tau_{\rm DP}+1}\left(\sigma_{\rm int}^{\rm sH}+\sigma_{\rm sj}^{\rm sH}\right) , \\
P_{\rm sT}{}&{}={}&{}-S_0\frac{2m\alpha}{\hbar^2} \frac{\tau_s}{\left(\tau_s/\tau_{\rm DP}+1\right)^2}\left(\sigma_{\rm int}^{\rm sH}+\sigma_{\rm sj}^{\rm sH}\right) . \label{PsT1}
\eer
Here, $1/\tau_{\rm DP}=(2m\alpha/\hbar^2)^2D$ is the Dyakonov-Perel relaxation rate in the diffusive regime, with $D=v_F^2\tau/2$ the diffusion constant, whereas $\sigma_{\rm int}^{\rm sH}=(N_0e\hbar/4m)(2\tau/\tau_{\rm DP})$
and $\sigma_{\rm sj}^{\rm sH}=en\lambda^2/(4\hbar)$ are the intrinsic and side-jump spin Hall conductivity, respectively.  Note that for a 2DEG we have $N_0e\hbar/4m=e/8\pi\hbar$, giving the ``universal'' intrinsic spin Hall conductivity.\cite{sinova2004} Clearly, $P_{\rm sT}$ is in general nonlinear in temperature due to the $T$-dependence of the spin relaxation rates,
\be
\frac{1}{\tau_{\rm DP}}\sim \tau \sim \frac{1}{T}, \hspace*{10pt} \frac{1}{\tau_{s}}\sim \frac{1}{\tau} \sim T .
\ee
An experimental relevant setup would be an open circuit along $x$, i.e., along the direction where the thermal gradient is applied. Then, the electric field can be expressed by the thermal gradient as $E_x=S\nabla_x T$, where $S$ is the Seebeck coefficient. For a 2DEG with an energy-independent relaxation rate $S=S_0$. With the open circuit condition the thermal Edelstein polarization coefficient is given as a sum of electrical and thermal contributions, and reads
\be \label{PtEdef}
{\cal P}^t=S_0 P_{\rm sE} + P_{\rm sT} ,
\ee
which is shown in Fig.~\ref{PtE} for $\tau/\tau_s=0.01$.  The parameter $\tau_{s,r}/\tau_{{\rm DP},r}$,
the subscript $r$ indicating that the value of a given quantity is taken at room temperature,
gives the ratio of intrinsic to extrinsic spin-orbit coupling and is usually large.
As discussed in Appendix ~\ref{App_taus}, one typically expects 
$1\lesssim\tau_s/\tau_{\rm DP}\lesssim10^2$.
The thermal contribution $P_{\rm sT}$ is in general less relevant when intrinsic spin-orbit coupling dominates [Fig.~\ref{PtE}b)], and only gives a significant contribution above room temperature. According to Ref.~[\onlinecite{isasa2014}] this should correspond to a metal like Pt, whereas Fig.~(\ref{PtE}a) to one like Au. The temperature dependence is clearly nonlinear around room temperature.

\begin{figure}
\flushleft{a)}\\
\includegraphics{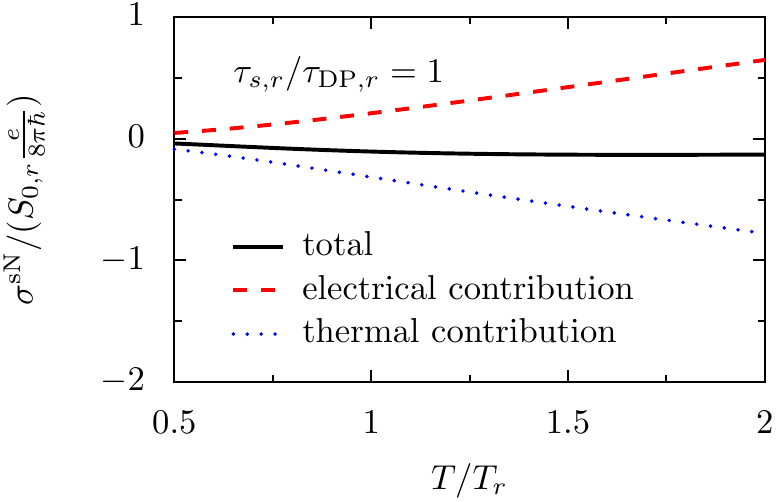}
\flushleft{a)}\\
\includegraphics{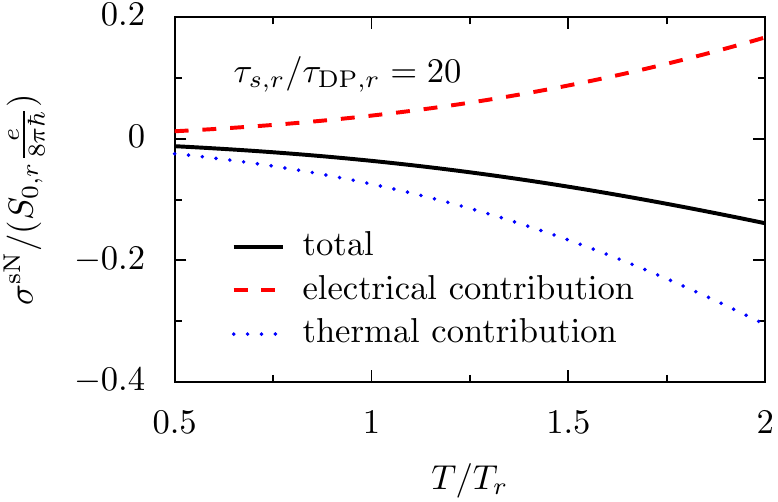}
\caption{Spin Nernst conductivity in units of the ``universal'' value of the intrinsic spin Hall conductivity times the Seebeck coefficient at room temperature, $S_{0,r}e/8\pi\hbar$, against $T/T_r$, with $\tau/\tau_s=0.01$. We show the electrical and thermal contribution separately; the parameters are chosen as $\tau_{s,r}/\tau_{{\rm DP},r}=1$ for a) and $\tau_{s,r}/\tau_{{\rm DP},r}=20$ for b). \label{sigmasN}}
\end{figure}

Analogously, we find for the spin current 
\ber
\sigma_{\rm sE}{}&{}={}&{}\frac{1}{\tau_s/\tau_{\rm DP}+1} \left(\sigma_{\rm int}^{\rm sH}+\sigma_{\rm sj}^{\rm sH}\right) , \\
\sigma_{\rm sT}{}&{}={}&{}-S_0\frac{2\tau_s/\tau_{\rm DP}+1}{\left(\tau_s/\tau_{\rm DP}+1\right)^2} \left(\sigma_{\rm int}^{\rm sH}+\sigma_{\rm sj}^{\rm sH}\right) ,
\eer
and for an open circuit condition the spin Nernst conductivity
\be
\label{snernst}
\sigma^{\rm sN}=S_0 \sigma_{\rm sE} + \sigma_{\rm sT} .
\ee
A plot of $\sigma^{\rm sN}$ versus temperature is shown in Fig.~\ref{sigmasN}, with $\tau/\tau_s=0.01$. The interplay of intrinsic and extrinsic spin-orbit coupling leads to a nonlinear temperature dependence, provided the intrinsic spin-orbit coupling dominates [Fig.~\ref{sigmasN}b)]. On the other hand, when intrinsic and extrinsic spin relaxation times are comparable [Fig.~\ref{sigmasN}a)] the spin Nernst conductivity is small since the thermal and the electrical contribution cancel each other. Indeed, for vanishing intrinsic spin-orbit coupling, the spin Nernst conductivity is zero for a 2DEG. 

Finally, we comment on the three-dimensional case. As can be seen in App.~\ref{App}, the quantitative change is rather small since only $\tau_s$ changes by a numerical pre-factor of $8/9$, while the other relevant quantities remain unchanged. We remark, however, that in 3D we encounter an energy-dependent density of states. In addition, the momentum relaxation rate is in general also energy-dependent. This manifests itself directly in the thermal part of the spin transport coefficients where we encounter the factors $\eta\equiv\epsilon_F N_0'/N_0$ and $\beta\equiv\epsilon_F \tau'/\tau$, namely the change in energy of the density of states and the momentum relaxation rate at the Fermi energy. Therefore, the relative weight between thermal and electrical contribution can be modified.  Note that in case of an open circuit along the thermal gradient, the electrical contribution is also modified by $\eta$ and $\beta$ since the Seebeck coefficient is then given by $S=S_0(1+\eta+\beta)$, as follows from the charge 
component of the Boltzmann equation.

\subsection{The ``clean'' limit}
\label{subsec_clean}

At room temperature, one enters the ``clean'' regime for $2\alpha p_F/\hbar>10^{-2}$eV. 
Under homogeneous conditions, Eq.~\eqref{Boltzmann} can be solved in this limit as well -- note
that Eq.~\eqref{conteq} is valid irrespective of the regime considered.  The procedure
is straightforward, yet lengthy and not particularly illuminating,\footnote{The idea is 
to write Eq.~\eqref{Boltzmann} in $4\times4$ matrix form, separating angular averaged quantities
from those which are not, and to solve it in linear response to the driving electric field
and temperature gradient.  See Ref.~[\onlinecite{raimondi2006}] for an explicit example.} therefore we simply give the results for the 2D case
when $2\alpha p_F \tau/\hbar^2\gg1$.  

The transport coefficients read
\ber
P_{\rm sE} 
&=&
-\frac{2m\alpha}{\hbar^2} 2\tau \left(\sigma_{\rm int}^{\rm sH}+\frac{\sigma_{\rm sj}^{\rm sH}}{2}\right) , 
\\
P_{\rm sT} 
&=&
S_0\frac{2m\alpha}{\hbar^2} 2\tau \frac{\sigma_{\rm sj}^{\rm sH}}{2} , \label{PsT2}
\\
\sigma_{\rm sE}
&=&
\frac{2\tau}{\tau_s} \left(\sigma_{\rm int}^{\rm sH}+\frac{\sigma_{\rm sj}^{\rm sH}}{2}\right), 
\\
\sigma_{\rm sT}
&=&
-S_0\frac{2\tau}{\tau_s} \sigma_{\rm sj}^{\rm sH},
\eer
where now $\sigma_{\rm int}^{\rm sH}=e/8\pi\hbar$.
From Eqs.~\eqref{PtEdef} and \eqref{snernst} it is immediate to see that the thermal Edelstein effect
is constant in $T$, whereas the spin Nernst is linear.  This overall simpler behavior is expected,
as in the ``clean'' limit $1/\tau_{DP}\rightarrow1/2\tau$, i.e., both the Dyakonov-Perel and the Elliott-Yafet 
relaxation rates are proportional to $T$.


\section{Conclusions}
\label{sec_conclusions}
We have explicitly considered the dynamical spin-orbit interaction of conduction electrons with phonons, 
which gives rise to dynamical Elliott-Yafet spin relaxation and side-jump mechanism.  The focus has been on the
high-temperature regime $T>T_D$.  Symmetric, Mott-like formulas for the (thermal) Edelstein and spin Hall and Nernst coefficients have been derived.  
The temperature-dependence of the spin transport coefficients was shown to be nontrivially affected by the competition between extrinsic and intrinsic 
spin-orbit coupling mechanisms, the origin lying in the mixing of the spin relaxation times $\tau_{\rm DP}$ and $\tau_s$.  
In the diffusive regime the latter have different temperature dependences, which ultimately causes the thermal Edelstein and spin Nernst effect to exhibit
a nonlinear $T$-behavior.  The nonlinearity is in general stronger for the thermal Edelstein effect, and,
especially in the spin Nernst case, it becomes weaker with decreasing intrinsic spin-orbit coupling strength.

\begin{acknowledgments}
We acknowledge stimulating discussions with R. Raimondi, as well as financial support from the German Research Foundation (DFG) through TRR 80, and from the CEA through the DSM-Energy Program (project E112-7-Meso-Therm-DSM).
\end{acknowledgments}

\appendix
\section{General Expressions for the spin polarization/current}
\label{App}
This appendix shows more general expressions for $P_{\rm sE}, P_{\rm sT}, \sigma_{\rm sE}$, and $\sigma_{\rm sT}$, 
valid at finite frequency for both 2D and 3D systems.
The transport coefficients are obtained by the Sommerfeld expansion of Eqs.~(\ref{Mott1})--(\ref{Mott2}). This implies that all quantities appearing below are evaluated at the Fermi energy unless otherwise specified. 

\begin{figure}[b]
\includegraphics[width=8cm]{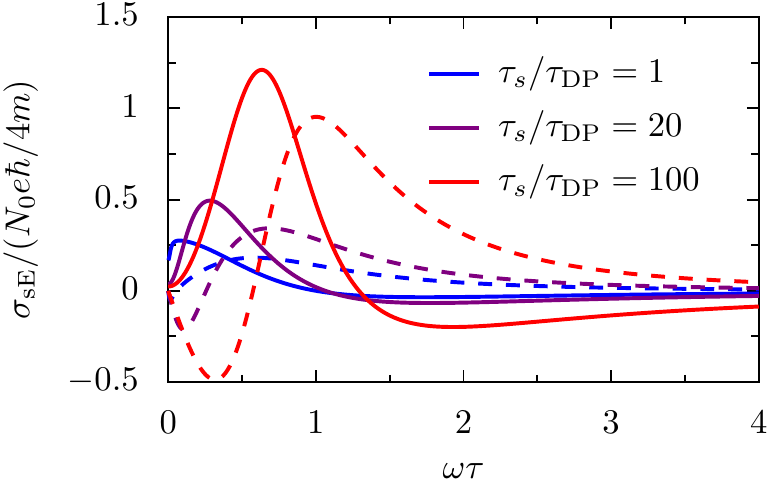}
\caption{The spin Hall conductivity in 3D in units of $N_0e\hbar/4m$ for various ratios of $\tau_s/\tau_{\rm DP}$ vs.\ $\omega\tau$, separated into its real part (solid lines) and its imaginary part (dashed lines). The extrinsic spin-orbit strength is chosen such that $\tau/\tau_s=0.01$. \label{sigmasE}}
\end{figure}

The dynamical Edelstein coefficient and the spin Hall conductivity are given as follows:
\ber
P_{\rm sE}{}&{}={}&{}-\frac{2m\alpha}{\hbar^2}\left[\frac{\tau_s}{2\tau_s/\tau_{\rm DP}+d(1-i\omega\tau_s)(1-i\omega\tau)}\right] \nonumber \\
{}&{}&{} \hspace*{40pt} \times \left[\frac{2\sigma_{\rm int}^{\rm sH} + d\sigma_{\rm sj}^{\rm sH} (1-i\omega\tau)}{1-i\omega\tau} \right] , \\
\sigma_{\rm sE}{}&{}={}&{}\left[ \frac{1-i\omega\tau_s}{2\tau_s/\tau_{\rm DP}+d(1-i\omega\tau_s)(1-i\omega\tau)} \right] \nonumber \\
{}&{}&{} \hspace*{40pt} \times \left[ \frac{2\sigma_{\rm int}^{\rm sH} + d\sigma_{\rm sj}^{\rm sH} (1-i\omega\tau)}{1-i\omega\tau} \right] .
\eer
Here, the form of $\sigma_{\rm int}^{\rm sH}$ and $\sigma_{\rm sj}^{\rm sH}$ [see Eq.~(\ref{PsT1})] remains unchanged in 3D and the Dyakonov-Perel relaxation rate remains exactly as it is in 2D, i.e., the diffusion constant which there appears is the 2D one.  Only the Elliott-Yafet relaxation rate exhibits a pre-factor of $8/9$ compared to $\tau_s$ in 2D.  
A plot of the spin Hall conductivity $\sigma_{\rm sE}(\omega)$ is shown in Fig.~\ref{sigmasE}.

The thermal contribution is now obtained by Eqs.~(\ref{Mott3}) and (\ref{Mott4}). Since the resulting equations are rather cumbersome, we just show formulas for the static case, $\omega=0$. We find
\ber
P_{\rm sT}{}&{}={}&{} - S_0 \frac{2m\alpha}{\hbar^2} \frac{\tau_s}{(2\tau_s/\tau_{\rm DP}+d)^2} \nonumber \\ 
{}&{}&{}\times\left\{ 2\sigma_{\rm int}^{\rm sH} \left[d-\eta\left(\frac{2\tau_s}{\tau_{\rm DP}+d}\right)-\beta\left(\frac{2\tau_s}{\tau_{\rm DP}+3d}\right) \right]\right. \nonumber \\
{}&{}&{}+\left. d\sigma_{\rm sj}^{\rm sH} \left[d-\eta\left(\frac{2\tau_s}{\tau_{\rm DP}+d}\right)+\beta\left(\frac{2\tau_s}{\tau_{\rm DP}-d}\right) \right]\right\} , \nonumber \\ \\
\sigma_{\rm sT}{}&{}={}&{} -S_0 \frac{1}{(2\tau_s/\tau_{\rm DP}+d)^2}  \nonumber \\
{}&{}&{}\times\left\{ 2\sigma_{\rm int}^{\rm sH} \left[ \frac{4\tau_s}{\tau_{\rm DP}}+d+\eta\left( \frac{2\tau_s}{\tau_{\rm DP}}+d \right) +2d\beta \right] \right. \nonumber \\
{}&{}&{}\hspace*{20pt}+\left. d\sigma_{\rm sj}^{\rm sH} \left[\frac{4\tau_s}{\tau_{\rm DP}} +d +\eta \left( \frac{2\tau_s}{\tau_{\rm DP}} +d \right)\right]\right\} .
\eer
Note that here the energy derivative of the density of states (momentum relaxation rate) at the Fermi energy comes into play by $\eta= \epsilon_F N_0'/N_0$ ($\beta= \epsilon_F \tau'/\tau$) which does have an influence on the thermal contribution to the spin Nernst conductivity and the spin polarization in case of an open circuit. But we remark that also the electrical contribution, $S\sigma_{\rm sE}$, is affected by $\eta$ and $\beta$ since the Seebeck coefficient then reads $S=S_0(1+\eta+\beta)$.

\section{On the ratio $\tau_s/\tau_{\rm DP}$}
\label{App_taus}
We estimate the size of the ratio $\tau_s/\tau_{\rm DP}$, defining the relative importance of extrinsic and intrinsic spin-orbit coupling.  The general form of the Dyakonov-Perel relaxation rate, valid from the ``clean'' to the ``dirty'' regime, reads
\be
\frac{1}{\tau_{\rm DP}}=\frac{1}{2\tau}\frac{(2\alpha p_F\tau/\hbar^2)^2}{(2\alpha p_F\tau/\hbar^2)^{2}+1},
\ee
where $2\alpha p_F/\hbar$ is the spin-orbit splitting.
Therefore
\be
\label{tauratio1}
\frac{\tau_{s}}{\tau_{\rm DP}} = \frac{1}{2}\frac{(2\alpha p_F\tau/\hbar^2)^2}{(2\alpha p_F\tau/\hbar^2)^{2}+1}
\left(\frac{2\hbar}{\lambda p_F}\right)^4.
\ee
In doped semiconductors one typically finds $10^{-2}\lesssim\lambda/\lambda_F\lesssim1$.
\cite{Handbook,winklerbook,vignale2010}  
Though there is yet no theory capable of estimating $\lambda$ in a metal,\cite{vignale2010}
one can argue that, since the spin-orbit energy is small compared to the Fermi one, the relation 
$\lambda/\lambda_F\ll1$ will hold in a metallic film.  Taking $\lambda/\lambda_F\approx10^{-1}$ yields $1\lesssim\tau_s/\tau_{\rm DP}\lesssim10^2$.  The lower value is valid in ``dirty'' (or with weak intrinsic spin-orbit) 
samples, $2\alpha p_F\tau/\hbar^2\lesssim10^{-2}$, the upper one in ``clean'' (or with strong intrinsic spin-orbit)
ones, $2\alpha p_F\tau/\hbar^2>1$.

\bibliography{paper_highT_biblio}

\begin{thebibliography}{57}%
\makeatletter
\providecommand \@ifxundefined [1]{%
 \@ifx{#1\undefined}
}%
\providecommand \@ifnum [1]{%
 \ifnum #1\expandafter \@firstoftwo
 \else \expandafter \@secondoftwo
 \fi
}%
\providecommand \@ifx [1]{%
 \ifx #1\expandafter \@firstoftwo
 \else \expandafter \@secondoftwo
 \fi
}%
\providecommand \natexlab [1]{#1}%
\providecommand \enquote  [1]{``#1''}%
\providecommand \bibnamefont  [1]{#1}%
\providecommand \bibfnamefont [1]{#1}%
\providecommand \citenamefont [1]{#1}%
\providecommand \href@noop [0]{\@secondoftwo}%
\providecommand \href [0]{\begingroup \@sanitize@url \@href}%
\providecommand \@href[1]{\@@startlink{#1}\@@href}%
\providecommand \@@href[1]{\endgroup#1\@@endlink}%
\providecommand \@sanitize@url [0]{\catcode `\\12\catcode `\$12\catcode
  `\&12\catcode `\#12\catcode `\^12\catcode `\_12\catcode `\%12\relax}%
\providecommand \@@startlink[1]{}%
\providecommand \@@endlink[0]{}%
\providecommand \url  [0]{\begingroup\@sanitize@url \@url }%
\providecommand \@url [1]{\endgroup\@href {#1}{\urlprefix }}%
\providecommand \urlprefix  [0]{URL }%
\providecommand \Eprint [0]{\href }%
\providecommand \doibase [0]{http://dx.doi.org/}%
\providecommand \selectlanguage [0]{\@gobble}%
\providecommand \bibinfo  [0]{\@secondoftwo}%
\providecommand \bibfield  [0]{\@secondoftwo}%
\providecommand \translation [1]{[#1]}%
\providecommand \BibitemOpen [0]{}%
\providecommand \bibitemStop [0]{}%
\providecommand \bibitemNoStop [0]{.\EOS\space}%
\providecommand \EOS [0]{\spacefactor3000\relax}%
\providecommand \BibitemShut  [1]{\csname bibitem#1\endcsname}%
\let\auto@bib@innerbib\@empty
\bibitem [{\citenamefont {Bauer}\ \emph {et~al.}(2012)\citenamefont {Bauer},
  \citenamefont {Saitoh},\ and\ \citenamefont {van Wees}}]{bauer2012}%
  \BibitemOpen
  \bibfield  {author} {\bibinfo {author} {\bibfnamefont {G.~E.~W.}\
  \bibnamefont {Bauer}}, \bibinfo {author} {\bibfnamefont {E.}~\bibnamefont
  {Saitoh}}, \ and\ \bibinfo {author} {\bibfnamefont {B.~J.}\ \bibnamefont {van
  Wees}},\ }\href@noop {} {\bibfield  {journal} {\bibinfo  {journal} {Nature
  Mater.}\ }\textbf {\bibinfo {volume} {11}},\ \bibinfo {pages} {391} (\bibinfo
  {year} {2012})}\BibitemShut {NoStop}%
\bibitem [{\citenamefont {Tauber}\ \emph {et~al.}(2012)\citenamefont {Tauber},
  \citenamefont {Gradhand}, \citenamefont {Fedorov},\ and\ \citenamefont
  {Mertig}}]{tauber2012}%
  \BibitemOpen
  \bibfield  {author} {\bibinfo {author} {\bibfnamefont {K.}~\bibnamefont
  {Tauber}}, \bibinfo {author} {\bibfnamefont {M.}~\bibnamefont {Gradhand}},
  \bibinfo {author} {\bibfnamefont {D.~V.}\ \bibnamefont {Fedorov}}, \ and\
  \bibinfo {author} {\bibfnamefont {I.}~\bibnamefont {Mertig}},\ }\href@noop {}
  {\bibfield  {journal} {\bibinfo  {journal} {Phys. Rev. Lett.}\ }\textbf
  {\bibinfo {volume} {109}},\ \bibinfo {pages} {026601} (\bibinfo {year}
  {2012})}\BibitemShut {NoStop}%
\bibitem [{\citenamefont {Borge}\ \emph {et~al.}(2013)\citenamefont {Borge},
  \citenamefont {Gorini},\ and\ \citenamefont {Raimondi}}]{borge2013}%
  \BibitemOpen
  \bibfield  {author} {\bibinfo {author} {\bibfnamefont {J.}~\bibnamefont
  {Borge}}, \bibinfo {author} {\bibfnamefont {C.}~\bibnamefont {Gorini}}, \
  and\ \bibinfo {author} {\bibfnamefont {R.}~\bibnamefont {Raimondi}},\ }\href
  {\doibase 10.1103/PhysRevB.87.085309} {\bibfield  {journal} {\bibinfo
  {journal} {Phys. Rev. B}\ }\textbf {\bibinfo {volume} {87}},\ \bibinfo
  {pages} {085309} (\bibinfo {year} {2013})}\BibitemShut {NoStop}%
\bibitem [{\citenamefont {Wang}\ and\ \citenamefont {Pang}(2010)}]{pang2010}%
  \BibitemOpen
  \bibfield  {author} {\bibinfo {author} {\bibfnamefont {C.~M.}\ \bibnamefont
  {Wang}}\ and\ \bibinfo {author} {\bibfnamefont {M.~Q.}\ \bibnamefont
  {Pang}},\ }\href@noop {} {\bibfield  {journal} {\bibinfo  {journal} {Solid
  State Commun.}\ }\textbf {\bibinfo {volume} {150}},\ \bibinfo {pages} {1509}
  (\bibinfo {year} {2010})}\BibitemShut {NoStop}%
\bibitem [{\citenamefont {Dyrda\l}\ \emph {et~al.}(2013)\citenamefont
  {Dyrda\l}, \citenamefont {Inglot}, \citenamefont {Dugaev},\ and\
  \citenamefont {Barna\'{s}}}]{dyrdal2013}%
  \BibitemOpen
  \bibfield  {author} {\bibinfo {author} {\bibfnamefont {A.}~\bibnamefont
  {Dyrda\l}}, \bibinfo {author} {\bibfnamefont {M.}~\bibnamefont {Inglot}},
  \bibinfo {author} {\bibfnamefont {V.~K.}\ \bibnamefont {Dugaev}}, \ and\
  \bibinfo {author} {\bibfnamefont {J.}~\bibnamefont {Barna\'{s}}},\
  }\href@noop {} {\bibfield  {journal} {\bibinfo  {journal} {Phys. Rev. B}\
  }\textbf {\bibinfo {volume} {87}},\ \bibinfo {pages} {245309} (\bibinfo
  {year} {2013})}\BibitemShut {NoStop}%
\bibitem [{\citenamefont {Dyakonov}\ and\ \citenamefont
  {Perel}(1971)}]{dyakonov1971}%
  \BibitemOpen
  \bibfield  {author} {\bibinfo {author} {\bibfnamefont {M.~I.}\ \bibnamefont
  {Dyakonov}}\ and\ \bibinfo {author} {\bibfnamefont {V.~I.}\ \bibnamefont
  {Perel}},\ }\href@noop {} {\bibfield  {journal} {\bibinfo  {journal} {Phys.
  Lett. A}\ }\textbf {\bibinfo {volume} {35}},\ \bibinfo {pages} {459}
  (\bibinfo {year} {1971})}\BibitemShut {NoStop}%
\bibitem [{\citenamefont {Aronov}\ and\ \citenamefont
  {Lyanda-Geller}(1989)}]{aronov1989}%
  \BibitemOpen
  \bibfield  {author} {\bibinfo {author} {\bibfnamefont {A.~G.}\ \bibnamefont
  {Aronov}}\ and\ \bibinfo {author} {\bibfnamefont {Y.~B.}\ \bibnamefont
  {Lyanda-Geller}},\ }\href@noop {} {\bibfield  {journal} {\bibinfo  {journal}
  {JETP Lett.}\ }\textbf {\bibinfo {volume} {50}},\ \bibinfo {pages} {431}
  (\bibinfo {year} {1989})}\BibitemShut {NoStop}%
\bibitem [{\citenamefont {Edelstein}(1990)}]{edelstein1990}%
  \BibitemOpen
  \bibfield  {author} {\bibinfo {author} {\bibfnamefont {V.~M.}\ \bibnamefont
  {Edelstein}},\ }\href@noop {} {\bibfield  {journal} {\bibinfo  {journal}
  {Solid State Commun.}\ }\textbf {\bibinfo {volume} {73}},\ \bibinfo {pages}
  {233} (\bibinfo {year} {1990})}\BibitemShut {NoStop}%
\bibitem [{\citenamefont {Vila}\ \emph {et~al.}(2007)\citenamefont {Vila},
  \citenamefont {Kimura},\ and\ \citenamefont {Otani}}]{vila2007}%
  \BibitemOpen
  \bibfield  {author} {\bibinfo {author} {\bibfnamefont {L.}~\bibnamefont
  {Vila}}, \bibinfo {author} {\bibfnamefont {T.}~\bibnamefont {Kimura}}, \ and\
  \bibinfo {author} {\bibfnamefont {Y.}~\bibnamefont {Otani}},\ }\href@noop {}
  {\bibfield  {journal} {\bibinfo  {journal} {Phys. Rev. Lett.}\ }\textbf
  {\bibinfo {volume} {99}},\ \bibinfo {pages} {226604} (\bibinfo {year}
  {2007})}\BibitemShut {NoStop}%
\bibitem [{\citenamefont {Kimura}\ \emph {et~al.}(2007)\citenamefont {Kimura},
  \citenamefont {Otani}, \citenamefont {Sato}, \citenamefont {Takahashi},\ and\
  \citenamefont {Maekawa}}]{kimura2007}%
  \BibitemOpen
  \bibfield  {author} {\bibinfo {author} {\bibfnamefont {T.}~\bibnamefont
  {Kimura}}, \bibinfo {author} {\bibfnamefont {Y.}~\bibnamefont {Otani}},
  \bibinfo {author} {\bibfnamefont {T.}~\bibnamefont {Sato}}, \bibinfo {author}
  {\bibfnamefont {S.}~\bibnamefont {Takahashi}}, \ and\ \bibinfo {author}
  {\bibfnamefont {S.}~\bibnamefont {Maekawa}},\ }\href@noop {} {\bibfield
  {journal} {\bibinfo  {journal} {Phys. Rev. Lett.}\ }\textbf {\bibinfo
  {volume} {98}},\ \bibinfo {pages} {156601} (\bibinfo {year}
  {2007})}\BibitemShut {NoStop}%
\bibitem [{\citenamefont {Seki}\ \emph {et~al.}(2008)\citenamefont {Seki},
  \citenamefont {Hasegawa}, \citenamefont {Mitani}, \citenamefont {Takahashi},
  \citenamefont {Imamura}, \citenamefont {Maekawa}, \citenamefont {Nitta},\
  and\ \citenamefont {Takanashi}}]{seki2008}%
  \BibitemOpen
  \bibfield  {author} {\bibinfo {author} {\bibfnamefont {T.}~\bibnamefont
  {Seki}}, \bibinfo {author} {\bibfnamefont {Y.}~\bibnamefont {Hasegawa}},
  \bibinfo {author} {\bibfnamefont {S.}~\bibnamefont {Mitani}}, \bibinfo
  {author} {\bibfnamefont {S.}~\bibnamefont {Takahashi}}, \bibinfo {author}
  {\bibfnamefont {H.}~\bibnamefont {Imamura}}, \bibinfo {author} {\bibfnamefont
  {S.}~\bibnamefont {Maekawa}}, \bibinfo {author} {\bibfnamefont
  {J.}~\bibnamefont {Nitta}}, \ and\ \bibinfo {author} {\bibfnamefont
  {K.}~\bibnamefont {Takanashi}},\ }\href@noop {} {\bibfield  {journal}
  {\bibinfo  {journal} {Nature Mater.}\ }\textbf {\bibinfo {volume} {7}},\
  \bibinfo {pages} {125} (\bibinfo {year} {2008})}\BibitemShut {NoStop}%
\bibitem [{\citenamefont {Liu}\ \emph {et~al.}(2011)\citenamefont {Liu},
  \citenamefont {Moriyama}, \citenamefont {Ralph},\ and\ \citenamefont
  {Buhrman}}]{liu2011}%
  \BibitemOpen
  \bibfield  {author} {\bibinfo {author} {\bibfnamefont {L.}~\bibnamefont
  {Liu}}, \bibinfo {author} {\bibfnamefont {T.}~\bibnamefont {Moriyama}},
  \bibinfo {author} {\bibfnamefont {D.~C.}\ \bibnamefont {Ralph}}, \ and\
  \bibinfo {author} {\bibfnamefont {R.~A.}\ \bibnamefont {Buhrman}},\
  }\href@noop {} {\bibfield  {journal} {\bibinfo  {journal} {Phys. Rev. Lett.}\
  }\textbf {\bibinfo {volume} {106}},\ \bibinfo {pages} {036601} (\bibinfo
  {year} {2011})}\BibitemShut {NoStop}%
\bibitem [{\citenamefont {Hahn}\ \emph {et~al.}(2013)\citenamefont {Hahn},
  \citenamefont {de~Loubens}, \citenamefont {Klein}, \citenamefont {Viretand},
  \citenamefont {Naletov},\ and\ \citenamefont {Youssef}}]{hahn2013}%
  \BibitemOpen
  \bibfield  {author} {\bibinfo {author} {\bibfnamefont {C.}~\bibnamefont
  {Hahn}}, \bibinfo {author} {\bibfnamefont {G.}~\bibnamefont {de~Loubens}},
  \bibinfo {author} {\bibfnamefont {O.}~\bibnamefont {Klein}}, \bibinfo
  {author} {\bibfnamefont {M.}~\bibnamefont {Viretand}}, \bibinfo {author}
  {\bibfnamefont {V.~V.}\ \bibnamefont {Naletov}}, \ and\ \bibinfo {author}
  {\bibfnamefont {J.~B.}\ \bibnamefont {Youssef}},\ }\href@noop {} {\bibfield
  {journal} {\bibinfo  {journal} {Phys. Rev. B}\ }\textbf {\bibinfo {volume}
  {87}},\ \bibinfo {pages} {174417} (\bibinfo {year} {2013})}\BibitemShut
  {NoStop}%
\bibitem [{\citenamefont {Kato}\ \emph {et~al.}(2004)\citenamefont {Kato},
  \citenamefont {Myers}, \citenamefont {Gossard},\ and\ \citenamefont
  {Awschalom}}]{kato2004}%
  \BibitemOpen
  \bibfield  {author} {\bibinfo {author} {\bibfnamefont {Y.~K.}\ \bibnamefont
  {Kato}}, \bibinfo {author} {\bibfnamefont {R.~C.}\ \bibnamefont {Myers}},
  \bibinfo {author} {\bibfnamefont {A.~C.}\ \bibnamefont {Gossard}}, \ and\
  \bibinfo {author} {\bibfnamefont {D.~D.}\ \bibnamefont {Awschalom}},\
  }\href@noop {} {\bibfield  {journal} {\bibinfo  {journal} {Science}\ }\textbf
  {\bibinfo {volume} {306}},\ \bibinfo {pages} {1910} (\bibinfo {year}
  {2004})}\BibitemShut {NoStop}%
\bibitem [{\citenamefont {Grimaldi}\ \emph {et~al.}(2006)\citenamefont
  {Grimaldi}, \citenamefont {Cappelluti},\ and\ \citenamefont
  {Marsiglio}}]{cappelluti2006}%
  \BibitemOpen
  \bibfield  {author} {\bibinfo {author} {\bibfnamefont {C.}~\bibnamefont
  {Grimaldi}}, \bibinfo {author} {\bibfnamefont {E.}~\bibnamefont
  {Cappelluti}}, \ and\ \bibinfo {author} {\bibfnamefont {F.}~\bibnamefont
  {Marsiglio}},\ }\href@noop {} {\bibfield  {journal} {\bibinfo  {journal}
  {Phys. Rev. Lett.}\ }\textbf {\bibinfo {volume} {97}},\ \bibinfo {pages}
  {066601} (\bibinfo {year} {2006})}\BibitemShut {NoStop}%
\bibitem [{\citenamefont {Bychkov}\ and\ \citenamefont
  {Rashba}(1984)}]{bychkov1984}%
  \BibitemOpen
  \bibfield  {author} {\bibinfo {author} {\bibfnamefont {Y.~A.}\ \bibnamefont
  {Bychkov}}\ and\ \bibinfo {author} {\bibfnamefont {E.~I.}\ \bibnamefont
  {Rashba}},\ }\href@noop {} {\bibfield  {journal} {\bibinfo  {journal} {J.
  Phys. C: Solid State Phys.}\ }\textbf {\bibinfo {volume} {17}},\ \bibinfo
  {pages} {6039} (\bibinfo {year} {1984})}\BibitemShut {NoStop}%
\bibitem [{\citenamefont {Gorini}\ \emph {et~al.}(2008)\citenamefont {Gorini},
  \citenamefont {Schwab}, \citenamefont {Dzierzawa},\ and\ \citenamefont
  {Raimondi}}]{gorini2008}%
  \BibitemOpen
  \bibfield  {author} {\bibinfo {author} {\bibfnamefont {C.}~\bibnamefont
  {Gorini}}, \bibinfo {author} {\bibfnamefont {P.}~\bibnamefont {Schwab}},
  \bibinfo {author} {\bibfnamefont {M.}~\bibnamefont {Dzierzawa}}, \ and\
  \bibinfo {author} {\bibfnamefont {R.}~\bibnamefont {Raimondi}},\ }\href@noop
  {} {\bibfield  {journal} {\bibinfo  {journal} {Phys. Rev. B}\ }\textbf
  {\bibinfo {volume} {78}},\ \bibinfo {pages} {125327} (\bibinfo {year}
  {2008})}\BibitemShut {NoStop}%
\bibitem [{\citenamefont {Raimondi}\ \emph {et~al.}(2012)\citenamefont
  {Raimondi}, \citenamefont {Schwab}, \citenamefont {Gorini},\ and\
  \citenamefont {Vignale}}]{raimondi2012}%
  \BibitemOpen
  \bibfield  {author} {\bibinfo {author} {\bibfnamefont {R.}~\bibnamefont
  {Raimondi}}, \bibinfo {author} {\bibfnamefont {P.}~\bibnamefont {Schwab}},
  \bibinfo {author} {\bibfnamefont {C.}~\bibnamefont {Gorini}}, \ and\ \bibinfo
  {author} {\bibfnamefont {G.}~\bibnamefont {Vignale}},\ }\href@noop {}
  {\bibfield  {journal} {\bibinfo  {journal} {Ann. Phys. (Berlin)}\ }\textbf
  {\bibinfo {volume} {524}},\ \bibinfo {pages} {153} (\bibinfo {year}
  {2012})}\BibitemShut {NoStop}%
\bibitem [{\citenamefont {Borge}\ \emph {et~al.}(2014)\citenamefont {Borge},
  \citenamefont {Gorini}, \citenamefont {Vignale},\ and\ \citenamefont
  {Raimondi}}]{borge2014}%
  \BibitemOpen
  \bibfield  {author} {\bibinfo {author} {\bibfnamefont {J.}~\bibnamefont
  {Borge}}, \bibinfo {author} {\bibfnamefont {C.}~\bibnamefont {Gorini}},
  \bibinfo {author} {\bibfnamefont {G.}~\bibnamefont {Vignale}}, \ and\
  \bibinfo {author} {\bibfnamefont {R.}~\bibnamefont {Raimondi}},\ }\href@noop
  {} {\bibfield  {journal} {\bibinfo  {journal} {Phys. Rev. B}\ }\textbf
  {\bibinfo {volume} {89}},\ \bibinfo {pages} {245443} (\bibinfo {year}
  {2014})}\BibitemShut {NoStop}%
\bibitem [{\citenamefont {Vignale}(2010)}]{vignale2010}%
  \BibitemOpen
  \bibfield  {author} {\bibinfo {author} {\bibfnamefont {G.}~\bibnamefont
  {Vignale}},\ }\href@noop {} {\bibfield  {journal} {\bibinfo  {journal} {J.
  Supercond. Nov. Magn.}\ }\textbf {\bibinfo {volume} {23}},\ \bibinfo {pages}
  {3} (\bibinfo {year} {2010})}\BibitemShut {NoStop}%
\bibitem [{\citenamefont {Niimi}\ \emph {et~al.}(2011)\citenamefont {Niimi},
  \citenamefont {Morota}, \citenamefont {Wei}, \citenamefont {Deranlot},
  \citenamefont {Basletic}, \citenamefont {Hamzic}, \citenamefont {Fert},\ and\
  \citenamefont {Otani}}]{niimi2011}%
  \BibitemOpen
  \bibfield  {author} {\bibinfo {author} {\bibfnamefont {Y.}~\bibnamefont
  {Niimi}}, \bibinfo {author} {\bibfnamefont {M.}~\bibnamefont {Morota}},
  \bibinfo {author} {\bibfnamefont {D.}~\bibnamefont {Wei}}, \bibinfo {author}
  {\bibfnamefont {C.}~\bibnamefont {Deranlot}}, \bibinfo {author}
  {\bibfnamefont {M.}~\bibnamefont {Basletic}}, \bibinfo {author}
  {\bibfnamefont {A.}~\bibnamefont {Hamzic}}, \bibinfo {author} {\bibfnamefont
  {A.}~\bibnamefont {Fert}}, \ and\ \bibinfo {author} {\bibfnamefont
  {Y.}~\bibnamefont {Otani}},\ }\href@noop {} {\bibfield  {journal} {\bibinfo
  {journal} {Phys. Rev. Lett.}\ }\textbf {\bibinfo {volume} {106}},\ \bibinfo
  {pages} {126601} (\bibinfo {year} {2011})}\BibitemShut {NoStop}%
\bibitem [{\citenamefont {Niimi}\ \emph {et~al.}(2014)\citenamefont {Niimi},
  \citenamefont {Suzuki}, \citenamefont {Kawanishi}, \citenamefont {Omori},
  \citenamefont {Valet}, \citenamefont {Fert},\ and\ \citenamefont
  {Otani}}]{niimi2014}%
  \BibitemOpen
  \bibfield  {author} {\bibinfo {author} {\bibfnamefont {Y.}~\bibnamefont
  {Niimi}}, \bibinfo {author} {\bibfnamefont {H.}~\bibnamefont {Suzuki}},
  \bibinfo {author} {\bibfnamefont {Y.}~\bibnamefont {Kawanishi}}, \bibinfo
  {author} {\bibfnamefont {Y.}~\bibnamefont {Omori}}, \bibinfo {author}
  {\bibfnamefont {T.}~\bibnamefont {Valet}}, \bibinfo {author} {\bibfnamefont
  {A.}~\bibnamefont {Fert}}, \ and\ \bibinfo {author} {\bibfnamefont
  {Y.}~\bibnamefont {Otani}},\ }\href@noop {} {\bibfield  {journal} {\bibinfo
  {journal} {Phys. Rev. B}\ }\textbf {\bibinfo {volume} {89}},\ \bibinfo
  {pages} {054401} (\bibinfo {year} {2014})}\BibitemShut {NoStop}%
\bibitem [{\citenamefont {Isasa}\ \emph {et~al.}(2014)\citenamefont {Isasa},
  \citenamefont {Villamor}, \citenamefont {Hueso}, \citenamefont {Gradhand},\
  and\ \citenamefont {Casanova}}]{isasa2014}%
  \BibitemOpen
  \bibfield  {author} {\bibinfo {author} {\bibfnamefont {M.}~\bibnamefont
  {Isasa}}, \bibinfo {author} {\bibfnamefont {E.}~\bibnamefont {Villamor}},
  \bibinfo {author} {\bibfnamefont {L.~E.}\ \bibnamefont {Hueso}}, \bibinfo
  {author} {\bibfnamefont {M.}~\bibnamefont {Gradhand}}, \ and\ \bibinfo
  {author} {\bibfnamefont {F.}~\bibnamefont {Casanova}},\ }\href@noop {}
  {\bibfield  {journal} {\bibinfo  {journal} {arXiv:}\ }\textbf {\bibinfo
  {volume} {1407.4770}} (\bibinfo {year} {2014})}\BibitemShut {NoStop}%
\bibitem [{\citenamefont {Ziman}(1972)}]{zimanbook1}%
  \BibitemOpen
  \bibfield  {author} {\bibinfo {author} {\bibfnamefont {J.~M.}\ \bibnamefont
  {Ziman}},\ }\href@noop {} {\emph {\bibinfo {title} {Principles of the Theory
  of Solids}}}\ (\bibinfo  {publisher} {Cambridge University Press},\ \bibinfo
  {year} {1972})\BibitemShut {NoStop}%
\bibitem [{\citenamefont {Shikin}\ \emph {et~al.}(2008)\citenamefont {Shikin},
  \citenamefont {Varykhalov}, \citenamefont {Prudnikova}, \citenamefont
  {Usachov}, \citenamefont {Adamchuk}, \citenamefont {Yamada}, \citenamefont
  {Riley},\ and\ \citenamefont {Rader}}]{shikin2008}%
  \BibitemOpen
  \bibfield  {author} {\bibinfo {author} {\bibfnamefont {A.~M.}\ \bibnamefont
  {Shikin}}, \bibinfo {author} {\bibfnamefont {A.}~\bibnamefont {Varykhalov}},
  \bibinfo {author} {\bibfnamefont {G.~V.}\ \bibnamefont {Prudnikova}},
  \bibinfo {author} {\bibfnamefont {D.}~\bibnamefont {Usachov}}, \bibinfo
  {author} {\bibfnamefont {V.~K.}\ \bibnamefont {Adamchuk}}, \bibinfo {author}
  {\bibfnamefont {Y.}~\bibnamefont {Yamada}}, \bibinfo {author} {\bibfnamefont
  {J.~D.}\ \bibnamefont {Riley}}, \ and\ \bibinfo {author} {\bibfnamefont
  {O.}~\bibnamefont {Rader}},\ }\href@noop {} {\bibfield  {journal} {\bibinfo
  {journal} {Phys. Rev. Lett.}\ }\textbf {\bibinfo {volume} {100}},\ \bibinfo
  {pages} {057601} (\bibinfo {year} {2008})}\BibitemShut {NoStop}%
\bibitem [{\citenamefont {Varykhalov}\ \emph {et~al.}(2008)\citenamefont
  {Varykhalov}, \citenamefont {S\'{a}nchez-Barriga}, \citenamefont {Shikin},
  \citenamefont {Gudat}, \citenamefont {Eberhardt},\ and\ \citenamefont
  {Rader}}]{varykhalov2008}%
  \BibitemOpen
  \bibfield  {author} {\bibinfo {author} {\bibfnamefont {A.}~\bibnamefont
  {Varykhalov}}, \bibinfo {author} {\bibfnamefont {J.}~\bibnamefont
  {S\'{a}nchez-Barriga}}, \bibinfo {author} {\bibfnamefont {A.~M.}\
  \bibnamefont {Shikin}}, \bibinfo {author} {\bibfnamefont {W.}~\bibnamefont
  {Gudat}}, \bibinfo {author} {\bibfnamefont {W.}~\bibnamefont {Eberhardt}}, \
  and\ \bibinfo {author} {\bibfnamefont {O.}~\bibnamefont {Rader}},\
  }\href@noop {} {\bibfield  {journal} {\bibinfo  {journal} {Phys. Rev. Lett.}\
  }\textbf {\bibinfo {volume} {101}},\ \bibinfo {pages} {256601} (\bibinfo
  {year} {2008})}\BibitemShut {NoStop}%
\bibitem [{\citenamefont {Rybkin}\ \emph {et~al.}(2010)\citenamefont {Rybkin},
  \citenamefont {Shikin}, \citenamefont {Adamchuk}, \citenamefont {Marchenko},
  \citenamefont {Biswas}, \citenamefont {Varykhalov},\ and\ \citenamefont
  {Rader}}]{rybkin2010}%
  \BibitemOpen
  \bibfield  {author} {\bibinfo {author} {\bibfnamefont {A.~G.}\ \bibnamefont
  {Rybkin}}, \bibinfo {author} {\bibfnamefont {A.~M.}\ \bibnamefont {Shikin}},
  \bibinfo {author} {\bibfnamefont {V.~K.}\ \bibnamefont {Adamchuk}}, \bibinfo
  {author} {\bibfnamefont {D.}~\bibnamefont {Marchenko}}, \bibinfo {author}
  {\bibfnamefont {C.}~\bibnamefont {Biswas}}, \bibinfo {author} {\bibfnamefont
  {A.}~\bibnamefont {Varykhalov}}, \ and\ \bibinfo {author} {\bibfnamefont
  {O.}~\bibnamefont {Rader}},\ }\href@noop {} {\bibfield  {journal} {\bibinfo
  {journal} {Phys. Rev. B}\ }\textbf {\bibinfo {volume} {82}},\ \bibinfo
  {pages} {233403} (\bibinfo {year} {2010})}\BibitemShut {NoStop}%
\bibitem [{\citenamefont {Hortamani}\ and\ \citenamefont
  {Wiesendanger}(2012)}]{hortamani2012}%
  \BibitemOpen
  \bibfield  {author} {\bibinfo {author} {\bibfnamefont {M.}~\bibnamefont
  {Hortamani}}\ and\ \bibinfo {author} {\bibfnamefont {R.}~\bibnamefont
  {Wiesendanger}},\ }\href@noop {} {\bibfield  {journal} {\bibinfo  {journal}
  {Phys. Rev. B}\ }\textbf {\bibinfo {volume} {86}},\ \bibinfo {pages} {235437}
  (\bibinfo {year} {2012})}\BibitemShut {NoStop}%
\bibitem [{\citenamefont {Winkler}(2003)}]{winklerbook}%
  \BibitemOpen
  \bibfield  {author} {\bibinfo {author} {\bibfnamefont {R.}~\bibnamefont
  {Winkler}},\ }\href@noop {} {\emph {\bibinfo {title} {Spin-Orbit Coupling
  Effects in Two-Dimensional Electron and Hole Systems}}}\ (\bibinfo
  {publisher} {Springer},\ \bibinfo {year} {2003})\BibitemShut {NoStop}%
\bibitem [{\citenamefont {Engel}\ \emph {et~al.}(2007)\citenamefont {Engel},
  \citenamefont {Rashba},\ and\ \citenamefont {Halperin}}]{Handbook}%
  \BibitemOpen
  \bibfield  {author} {\bibinfo {author} {\bibfnamefont {H.-A.}\ \bibnamefont
  {Engel}}, \bibinfo {author} {\bibfnamefont {E.~I.}\ \bibnamefont {Rashba}}, \
  and\ \bibinfo {author} {\bibfnamefont {B.~I.}\ \bibnamefont {Halperin}},\
  }in\ \href@noop {} {\emph {\bibinfo {booktitle} {Handbook of Magnetism and
  Advanced Magnetic Materials}}},\ Vol.~\bibinfo {volume} {V},\ \bibinfo
  {editor} {edited by\ \bibinfo {editor} {\bibfnamefont {H.}~\bibnamefont
  {Kronm\"uller}}\ and\ \bibinfo {editor} {\bibfnamefont {S.}~\bibnamefont
  {Parkin}}}\ (\bibinfo  {publisher} {Wiley},\ \bibinfo {year} {2007})\ pp.\
  \bibinfo {pages} {2858--2877}\BibitemShut {NoStop}%
\bibitem [{\citenamefont {Valenzuela}\ and\ \citenamefont
  {Tinkham}(2006)}]{valenzuela2006}%
  \BibitemOpen
  \bibfield  {author} {\bibinfo {author} {\bibfnamefont {S.~O.}\ \bibnamefont
  {Valenzuela}}\ and\ \bibinfo {author} {\bibfnamefont {M.}~\bibnamefont
  {Tinkham}},\ }\href@noop {} {\bibfield  {journal} {\bibinfo  {journal}
  {Nature Mater.}\ }\textbf {\bibinfo {volume} {442}},\ \bibinfo {pages} {176}
  (\bibinfo {year} {2006})}\BibitemShut {NoStop}%
\bibitem [{\citenamefont {Pesin}\ and\ \citenamefont
  {MacDonald}(2012)}]{pesin2012}%
  \BibitemOpen
  \bibfield  {author} {\bibinfo {author} {\bibfnamefont {D.~A.}\ \bibnamefont
  {Pesin}}\ and\ \bibinfo {author} {\bibfnamefont {A.~H.}\ \bibnamefont
  {MacDonald}},\ }\href@noop {} {\bibfield  {journal} {\bibinfo  {journal}
  {Phys. Rev. B}\ }\textbf {\bibinfo {volume} {86}},\ \bibinfo {pages} {014416}
  (\bibinfo {year} {2012})}\BibitemShut {NoStop}%
\bibitem [{\citenamefont {Wang}\ \emph {et~al.}(2013)\citenamefont {Wang},
  \citenamefont {Xiao}, \citenamefont {Manchon},\ and\ \citenamefont
  {Maekawa}}]{wang2013}%
  \BibitemOpen
  \bibfield  {author} {\bibinfo {author} {\bibfnamefont {X.}~\bibnamefont
  {Wang}}, \bibinfo {author} {\bibfnamefont {J.}~\bibnamefont {Xiao}}, \bibinfo
  {author} {\bibfnamefont {A.}~\bibnamefont {Manchon}}, \ and\ \bibinfo
  {author} {\bibfnamefont {S.}~\bibnamefont {Maekawa}},\ }\href@noop {}
  {\bibfield  {journal} {\bibinfo  {journal} {Phys. Rev. B}\ }\textbf {\bibinfo
  {volume} {87}},\ \bibinfo {pages} {081407} (\bibinfo {year}
  {2013})}\BibitemShut {NoStop}%
\bibitem [{\citenamefont {Haney}\ \emph {et~al.}(2013)\citenamefont {Haney},
  \citenamefont {Lee}, \citenamefont {Lee}, \citenamefont {Manchon},\ and\
  \citenamefont {Stiles}}]{haney2013}%
  \BibitemOpen
  \bibfield  {author} {\bibinfo {author} {\bibfnamefont {P.~M.}\ \bibnamefont
  {Haney}}, \bibinfo {author} {\bibfnamefont {H.-W.}\ \bibnamefont {Lee}},
  \bibinfo {author} {\bibfnamefont {K.-J.}\ \bibnamefont {Lee}}, \bibinfo
  {author} {\bibfnamefont {A.}~\bibnamefont {Manchon}}, \ and\ \bibinfo
  {author} {\bibfnamefont {M.~D.}\ \bibnamefont {Stiles}},\ }\href@noop {}
  {\bibfield  {journal} {\bibinfo  {journal} {Phys. Rev. B}\ }\textbf {\bibinfo
  {volume} {87}},\ \bibinfo {pages} {174411} (\bibinfo {year}
  {2013})}\BibitemShut {NoStop}%
\bibitem [{\citenamefont {Gorini}\ \emph {et~al.}(2010)\citenamefont {Gorini},
  \citenamefont {Schwab}, \citenamefont {Raimondi},\ and\ \citenamefont
  {Shelankov}}]{gorini2010}%
  \BibitemOpen
  \bibfield  {author} {\bibinfo {author} {\bibfnamefont {C.}~\bibnamefont
  {Gorini}}, \bibinfo {author} {\bibfnamefont {P.}~\bibnamefont {Schwab}},
  \bibinfo {author} {\bibfnamefont {R.}~\bibnamefont {Raimondi}}, \ and\
  \bibinfo {author} {\bibfnamefont {A.~L.}\ \bibnamefont {Shelankov}},\
  }\href@noop {} {\bibfield  {journal} {\bibinfo  {journal} {Phys. Rev. B}\
  }\textbf {\bibinfo {volume} {82}},\ \bibinfo {pages} {195316} (\bibinfo
  {year} {2010})}\BibitemShut {NoStop}%
\bibitem [{\citenamefont {Tokatly}(2008)}]{tokatly2008}%
  \BibitemOpen
  \bibfield  {author} {\bibinfo {author} {\bibfnamefont {I.~V.}\ \bibnamefont
  {Tokatly}},\ }\href@noop {} {\bibfield  {journal} {\bibinfo  {journal} {Phys.
  Rev. Lett.}\ }\textbf {\bibinfo {volume} {101}},\ \bibinfo {pages} {106601}
  (\bibinfo {year} {2008})}\BibitemShut {NoStop}%
\bibitem [{\citenamefont {Gorini}\ \emph {et~al.}(2012)\citenamefont {Gorini},
  \citenamefont {Raimondi},\ and\ \citenamefont {Schwab}}]{gorini2012}%
  \BibitemOpen
  \bibfield  {author} {\bibinfo {author} {\bibfnamefont {C.}~\bibnamefont
  {Gorini}}, \bibinfo {author} {\bibfnamefont {R.}~\bibnamefont {Raimondi}}, \
  and\ \bibinfo {author} {\bibfnamefont {P.}~\bibnamefont {Schwab}},\
  }\href@noop {} {\bibfield  {journal} {\bibinfo  {journal} {Phys. Rev. Lett.}\
  }\textbf {\bibinfo {volume} {109}},\ \bibinfo {pages} {246604} (\bibinfo
  {year} {2012})}\BibitemShut {NoStop}%
\bibitem [{\citenamefont {Shen}\ \emph {et~al.}(2014)\citenamefont {Shen},
  \citenamefont {Vignale},\ and\ \citenamefont {Raimondi}}]{shen2014}%
  \BibitemOpen
  \bibfield  {author} {\bibinfo {author} {\bibfnamefont {K.}~\bibnamefont
  {Shen}}, \bibinfo {author} {\bibfnamefont {G.}~\bibnamefont {Vignale}}, \
  and\ \bibinfo {author} {\bibfnamefont {R.}~\bibnamefont {Raimondi}},\
  }\href@noop {} {\bibfield  {journal} {\bibinfo  {journal} {Phys. Rev. Lett.}\
  }\textbf {\bibinfo {volume} {112}},\ \bibinfo {pages} {096601} (\bibinfo
  {year} {2014})}\BibitemShut {NoStop}%
\bibitem [{Note1()}]{Note1}%
  \BibitemOpen
  \bibinfo {note} {Finer band structure details such as non-parabolicity can be
  included in the kinetic treatment to follow (Ref.~[\protect \rev@citealpnum
  {shytov2006}], though concerned with a semiconductor scenario, gives a taste
  of the technicalities involved.). However, since they are not central to our
  goals, we stick to a bare-bone model for simplicity's sake.}\BibitemShut
  {Stop}%
\bibitem [{\citenamefont {Hankiewicz}\ and\ \citenamefont
  {Vignale}(2008)}]{hankiewicz2008}%
  \BibitemOpen
  \bibfield  {author} {\bibinfo {author} {\bibfnamefont {E.~M.}\ \bibnamefont
  {Hankiewicz}}\ and\ \bibinfo {author} {\bibfnamefont {G.}~\bibnamefont
  {Vignale}},\ }\href@noop {} {\bibfield  {journal} {\bibinfo  {journal} {Phys.
  Rev. Lett.}\ }\textbf {\bibinfo {volume} {100}},\ \bibinfo {pages} {026602}
  (\bibinfo {year} {2008})}\BibitemShut {NoStop}%
\bibitem [{\citenamefont {Raimondi}\ and\ \citenamefont
  {Schwab}(2009)}]{raimondi2009}%
  \BibitemOpen
  \bibfield  {author} {\bibinfo {author} {\bibfnamefont {R.}~\bibnamefont
  {Raimondi}}\ and\ \bibinfo {author} {\bibfnamefont {P.}~\bibnamefont
  {Schwab}},\ }\href@noop {} {\bibfield  {journal} {\bibinfo  {journal} {EPL}\
  }\textbf {\bibinfo {volume} {87}},\ \bibinfo {pages} {37008} (\bibinfo {year}
  {2009})}\BibitemShut {NoStop}%
\bibitem [{\citenamefont {Sinitsyn}(2008)}]{sinitsyn2008}%
  \BibitemOpen
  \bibfield  {author} {\bibinfo {author} {\bibfnamefont {N.~A.}\ \bibnamefont
  {Sinitsyn}},\ }\href@noop {} {\bibfield  {journal} {\bibinfo  {journal} {J.
  Phys.: Condens. Matter}\ }\textbf {\bibinfo {volume} {20}},\ \bibinfo {pages}
  {023201} (\bibinfo {year} {2008})}\BibitemShut {NoStop}%
\bibitem [{\citenamefont {Culcer}\ \emph {et~al.}(2010)\citenamefont {Culcer},
  \citenamefont {Hankiewicz}, \citenamefont {Vignale},\ and\ \citenamefont
  {Winkler}}]{culcer2010}%
  \BibitemOpen
  \bibfield  {author} {\bibinfo {author} {\bibfnamefont {D.}~\bibnamefont
  {Culcer}}, \bibinfo {author} {\bibfnamefont {E.~M.}\ \bibnamefont
  {Hankiewicz}}, \bibinfo {author} {\bibfnamefont {G.}~\bibnamefont {Vignale}},
  \ and\ \bibinfo {author} {\bibfnamefont {R.}~\bibnamefont {Winkler}},\
  }\href@noop {} {\bibfield  {journal} {\bibinfo  {journal} {Phys. Rev. B}\
  }\textbf {\bibinfo {volume} {81}},\ \bibinfo {pages} {125332} (\bibinfo
  {year} {2010})}\BibitemShut {NoStop}%
\bibitem [{\citenamefont {Gorini}\ \emph {et~al.}(shed)\citenamefont {Gorini},
  \citenamefont {T\"{o}lle},\ and\ \citenamefont {Eckern}}]{skewfuture}%
  \BibitemOpen
  \bibfield  {author} {\bibinfo {author} {\bibfnamefont {C.}~\bibnamefont
  {Gorini}}, \bibinfo {author} {\bibfnamefont {S.}~\bibnamefont {T\"{o}lle}}, \
  and\ \bibinfo {author} {\bibfnamefont {U.}~\bibnamefont {Eckern}},\
  }\href@noop {} {\  (\bibinfo {year} {unpublished})}\BibitemShut {NoStop}%
\bibitem [{\citenamefont {Mathur}\ and\ \citenamefont
  {Stone}(1992)}]{mathur1992}%
  \BibitemOpen
  \bibfield  {author} {\bibinfo {author} {\bibfnamefont {H.}~\bibnamefont
  {Mathur}}\ and\ \bibinfo {author} {\bibfnamefont {A.~D.}\ \bibnamefont
  {Stone}},\ }\href@noop {} {\bibfield  {journal} {\bibinfo  {journal} {Phys.
  Rev. Lett.}\ }\textbf {\bibinfo {volume} {68}},\ \bibinfo {pages} {2964}
  (\bibinfo {year} {1992})}\BibitemShut {NoStop}%
\bibitem [{\citenamefont {Fr\"{o}hlich}\ and\ \citenamefont
  {Studer}(1993)}]{frohlich1993}%
  \BibitemOpen
  \bibfield  {author} {\bibinfo {author} {\bibfnamefont {J.}~\bibnamefont
  {Fr\"{o}hlich}}\ and\ \bibinfo {author} {\bibfnamefont {U.~M.}\ \bibnamefont
  {Studer}},\ }\href@noop {} {\bibfield  {journal} {\bibinfo  {journal} {Rev.
  Mod. Phys.}\ }\textbf {\bibinfo {volume} {65}},\ \bibinfo {pages} {733}
  (\bibinfo {year} {1993})}\BibitemShut {NoStop}%
\bibitem [{\citenamefont {Ganichev}\ \emph {et~al.}(2002)\citenamefont
  {Ganichev}, \citenamefont {Ivchenko}, \citenamefont {Bel'kov}, \citenamefont
  {Tarasenko}, \citenamefont {Sollinger}, \citenamefont {Weiss}, \citenamefont
  {Wegscheider},\ and\ \citenamefont {Prettl}}]{ganichev2002}%
  \BibitemOpen
  \bibfield  {author} {\bibinfo {author} {\bibfnamefont {S.~D.}\ \bibnamefont
  {Ganichev}}, \bibinfo {author} {\bibfnamefont {E.~L.}\ \bibnamefont
  {Ivchenko}}, \bibinfo {author} {\bibfnamefont {V.~V.}\ \bibnamefont
  {Bel'kov}}, \bibinfo {author} {\bibfnamefont {S.~A.}\ \bibnamefont
  {Tarasenko}}, \bibinfo {author} {\bibfnamefont {M.}~\bibnamefont
  {Sollinger}}, \bibinfo {author} {\bibfnamefont {D.}~\bibnamefont {Weiss}},
  \bibinfo {author} {\bibfnamefont {W.}~\bibnamefont {Wegscheider}}, \ and\
  \bibinfo {author} {\bibfnamefont {W.}~\bibnamefont {Prettl}},\ }\href@noop {}
  {\bibfield  {journal} {\bibinfo  {journal} {Nature}\ }\textbf {\bibinfo
  {volume} {417}},\ \bibinfo {pages} {153} (\bibinfo {year}
  {2002})}\BibitemShut {NoStop}%
\bibitem [{\citenamefont {S\'anchez}\ \emph {et~al.}(2013)\citenamefont
  {S\'anchez}, \citenamefont {Vila}, \citenamefont {Desfonds}, \citenamefont
  {Gambarelli}, \citenamefont {Attan\'e}, \citenamefont {Teresa}, \citenamefont
  {Mag\'en},\ and\ \citenamefont {Fert}}]{rojassanchez2013}%
  \BibitemOpen
  \bibfield  {author} {\bibinfo {author} {\bibfnamefont {J.~C.~R.}\
  \bibnamefont {S\'anchez}}, \bibinfo {author} {\bibfnamefont {L.}~\bibnamefont
  {Vila}}, \bibinfo {author} {\bibfnamefont {G.}~\bibnamefont {Desfonds}},
  \bibinfo {author} {\bibfnamefont {S.}~\bibnamefont {Gambarelli}}, \bibinfo
  {author} {\bibfnamefont {J.~P.}\ \bibnamefont {Attan\'e}}, \bibinfo {author}
  {\bibfnamefont {J.~M.~D.}\ \bibnamefont {Teresa}}, \bibinfo {author}
  {\bibfnamefont {C.}~\bibnamefont {Mag\'en}}, \ and\ \bibinfo {author}
  {\bibfnamefont {A.}~\bibnamefont {Fert}},\ }\href@noop {} {\bibfield
  {journal} {\bibinfo  {journal} {Nat. Commun.}\ }\textbf {\bibinfo {volume}
  {4}},\ \bibinfo {pages} {2944} (\bibinfo {year} {2013})}\BibitemShut
  {NoStop}%
\bibitem [{\citenamefont {Rammer}(1998)}]{rammer}%
  \BibitemOpen
  \bibfield  {author} {\bibinfo {author} {\bibfnamefont {J.}~\bibnamefont
  {Rammer}},\ }\href@noop {} {\emph {\bibinfo {title} {Quantum Transport
  Theory}}}\ (\bibinfo  {publisher} {Perseus Books},\ \bibinfo {year}
  {1998})\BibitemShut {NoStop}%
\bibitem [{\citenamefont {Rammer}(2011)}]{rammer2}%
  \BibitemOpen
  \bibfield  {author} {\bibinfo {author} {\bibfnamefont {J.}~\bibnamefont
  {Rammer}},\ }\href@noop {} {\emph {\bibinfo {title} {Quantum Field Theory of
  Non-equilibrium States}}}\ (\bibinfo  {publisher} {Cambridge University
  Press},\ \bibinfo {year} {2011})\BibitemShut {NoStop}%
\bibitem [{\citenamefont {Ashcroft}\ and\ \citenamefont
  {Mermin}(1976)}]{ashcroft}%
  \BibitemOpen
  \bibfield  {author} {\bibinfo {author} {\bibfnamefont {N.~W.}\ \bibnamefont
  {Ashcroft}}\ and\ \bibinfo {author} {\bibfnamefont {N.~D.}\ \bibnamefont
  {Mermin}},\ }\href@noop {} {\emph {\bibinfo {title} {Solid State Physics}}}\
  (\bibinfo  {publisher} {Brooks/Cole},\ \bibinfo {year} {1976})\BibitemShut
  {NoStop}%
\bibitem [{Note2()}]{Note2}%
  \BibitemOpen
  \bibinfo {note} {We remark that in three dimensions Eq.~(\ref {EYop}) is not
  the complete collision operator as obtained from the self-energy depicted in
  Fig.~\ref {Diagrams}: we have dropped terms where the momentum is not
  parallel to the Pauli vector. These contributions are negligible when
  investigating the spin transport quantities.}\BibitemShut {Stop}%
\bibitem [{\citenamefont {{Cutler}}\ and\ \citenamefont
  {{Mott}}(1969)}]{Mott1969}%
  \BibitemOpen
  \bibfield  {author} {\bibinfo {author} {\bibfnamefont {M.}~\bibnamefont
  {{Cutler}}}\ and\ \bibinfo {author} {\bibfnamefont {N.~F.}\ \bibnamefont
  {{Mott}}},\ }\href {\doibase 10.1103/PhysRev.181.1336} {\bibfield  {journal}
  {\bibinfo  {journal} {Phys. Rev.}\ }\textbf {\bibinfo {volume} {181}},\
  \bibinfo {pages} {1336} (\bibinfo {year} {1969})}\BibitemShut {NoStop}%
\bibitem [{\citenamefont {Sinova}\ \emph {et~al.}(2004)\citenamefont {Sinova},
  \citenamefont {Culcer}, \citenamefont {Niu}, \citenamefont {Sinitsyn},
  \citenamefont {Jungwirth},\ and\ \citenamefont {MacDonald}}]{sinova2004}%
  \BibitemOpen
  \bibfield  {author} {\bibinfo {author} {\bibfnamefont {J.}~\bibnamefont
  {Sinova}}, \bibinfo {author} {\bibfnamefont {D.}~\bibnamefont {Culcer}},
  \bibinfo {author} {\bibfnamefont {Q.}~\bibnamefont {Niu}}, \bibinfo {author}
  {\bibfnamefont {N.~A.}\ \bibnamefont {Sinitsyn}}, \bibinfo {author}
  {\bibfnamefont {T.}~\bibnamefont {Jungwirth}}, \ and\ \bibinfo {author}
  {\bibfnamefont {A.~H.}\ \bibnamefont {MacDonald}},\ }\href {\doibase
  10.1103/PhysRevLett.92.126603} {\bibfield  {journal} {\bibinfo  {journal}
  {Phys. Rev. Lett.}\ }\textbf {\bibinfo {volume} {92}},\ \bibinfo {pages}
  {126603} (\bibinfo {year} {2004})}\BibitemShut {NoStop}%
\bibitem [{Note3()}]{Note3}%
  \BibitemOpen
  \bibinfo {note} {The idea is to write Eq.~\protect \textup {\hbox
  {\mathsurround \z@ \protect \normalfont (\ignorespaces \ref
  {Boltzmann}\unskip \@@italiccorr )}} in $4\times 4$ matrix form, separating
  angular averaged quantities from those which are not, and to solve it in
  linear response to the driving electric field and temperature gradient. See
  Ref.~[\protect \rev@citealpnum {raimondi2006}] for an explicit
  example.}\BibitemShut {Stop}%
\bibitem [{\citenamefont {Shytov}\ \emph {et~al.}(2006)\citenamefont {Shytov},
  \citenamefont {Mishchenko}, \citenamefont {Engel},\ and\ \citenamefont
  {Halperin}}]{shytov2006}%
  \BibitemOpen
  \bibfield  {author} {\bibinfo {author} {\bibfnamefont {A.~V.}\ \bibnamefont
  {Shytov}}, \bibinfo {author} {\bibfnamefont {E.~G.}\ \bibnamefont
  {Mishchenko}}, \bibinfo {author} {\bibfnamefont {H.-A.}\ \bibnamefont
  {Engel}}, \ and\ \bibinfo {author} {\bibfnamefont {B.~I.}\ \bibnamefont
  {Halperin}},\ }\href@noop {} {\bibfield  {journal} {\bibinfo  {journal}
  {Phys. Rev. B}\ }\textbf {\bibinfo {volume} {73}},\ \bibinfo {pages} {075316}
  (\bibinfo {year} {2006})}\BibitemShut {NoStop}%
\bibitem [{\citenamefont {Raimondi}\ \emph {et~al.}(2006)\citenamefont
  {Raimondi}, \citenamefont {Gorini}, \citenamefont {Schwab},\ and\
  \citenamefont {Dzierzawa}}]{raimondi2006}%
  \BibitemOpen
  \bibfield  {author} {\bibinfo {author} {\bibfnamefont {R.}~\bibnamefont
  {Raimondi}}, \bibinfo {author} {\bibfnamefont {C.}~\bibnamefont {Gorini}},
  \bibinfo {author} {\bibfnamefont {P.}~\bibnamefont {Schwab}}, \ and\ \bibinfo
  {author} {\bibfnamefont {M.}~\bibnamefont {Dzierzawa}},\ }\href@noop {}
  {\bibfield  {journal} {\bibinfo  {journal} {Phys. Rev. B}\ }\textbf {\bibinfo
  {volume} {74}},\ \bibinfo {pages} {035340} (\bibinfo {year}
  {2006})}\BibitemShut {NoStop}%
\end{thebibliography}%
\end{document}